\documentclass[sigconf]{acmart}
\usepackage[linesnumbered,ruled,vlined]{algorithm2e}
\usepackage{amsmath}
\usepackage{framed}
\usepackage{booktabs}
\usepackage{subcaption}
\usepackage{todonotes}
\usepackage{comment}
\usepackage{multirow}
\usepackage{listings}
\usepackage{float}
\usepackage{bbding}

\newcommand{\tool}{{\textsc{ATTuzz}}\xspace}%

\begin{document}

\title{Better Pay Attention Whilst Fuzzing}

\author{Shunkai Zhu}
\email{shunkaiz@zju.edu.cn}
\affiliation{%
  \institution{Zhejiang University}
  \country{China}
}

\author{Jingyi Wang}
\authornote{Corresponding author.}
\email{wangjyee@zju.edu.cn}
\affiliation{%
  \institution{Zhejiang University}
  \country{China}
}

\author{Jun Sun}
\email{junsun@smu.edu.sg}
\affiliation{%
  \institution{Singapore Management University}
  \country{China}
}

\author{Jie Yang}
\email{y@zju.edu.cn}
\affiliation{%
  \institution{Zhejiang University}
  \country{China}
}

\author{Xingwei Lin}
\email{xwlin.roy@gmail.com}
\affiliation{%
  \institution{Ant Group}
  \country{China}
}

\author{Liyi Zhang}
\email{l392zhan@uwaterloo.ca}
\affiliation{%
  \institution{Zhejiang University}
  \country{China}
}

\author{Peng Cheng}
\email{lunarheart@zju.edu.cn}
\affiliation{%
  \institution{Zhejiang University}
  \country{China}
}
\renewcommand{\shortauthors}{Trovato and Tobin, et al.}

\begin{abstract}
Fuzzing is one of the prevailing methods for vulnerability detection. However, even state-of-the-art fuzzing methods become ineffective after some period of time, i.e., the coverage hardly improves as existing methods are ineffective to focus the attention of fuzzing on covering the hard-to-trigger program paths. In other words, they cannot generate inputs that can break the bottleneck due to the fundamental difficulty in capturing the complex relations between the test inputs and program coverage. 
% There are two main reasons. First, 
In particular, existing fuzzers suffer from the following main limitations: 1) lacking an overall analysis of the program to identify the most ``rewarding'' seeds, and 2) lacking an effective mutation strategy which could continuously select and mutates the more relevant ``bytes'' of the seeds. 
% The fundamental challenge is the difficulty to capture the complex relations between the test inputs and their capability of covering program branches. 
% The fundamental reason is that there are often complex relationship between test inputs and their capability of covering certain program branches, which renders approaches that approximate such relationship with simple models ineffective over time.

In this work, we propose an approach called \tool to address these two issues systematically. First, we propose a lightweight dynamic analysis technique which estimates the ``reward" of covering each basic block and selects the most rewarding seeds accordingly. Second, we mutate the selected seeds according to a neural network model which predicts whether a certain ``rewarding'' block will be covered given certain mutation on certain bytes of a seed. The model is a deep learning model equipped with attention mechanism which is learned and updated periodically whilst fuzzing. Our evaluation shows that \tool significantly outperforms 5 state-of-the-art grey-box fuzzers on 13 popular real-world programs at achieving higher edge coverage and finding new bugs. In particular, \tool achieved 2X edge coverage and 4X bugs detected than AFL over 24-hour runs. Moreover, \tool persistently improves the edge coverage in the long run, i.e., achieving 50\% more coverage than AFL in 5 days. 

\end{abstract}

\maketitle
\section{Introduction}

Fuzzing has become one of the prevailing methods for vulnerability detection. It works by generating ``random'' test inputs to execute the target program, aiming to trigger potential security vulnerabilities~\cite{miller1995fuzz,embleton2006sidewinder}. Thanks to its simple and easy-to-apply concept, fuzzing has been widely adopted to test real-world programs~\cite{petsios2017slowfuzz,8233151,libFuzzer,AFL,serebryany2017oss}. 
% However, typical fuzzers either blindly select seed files or mutate seed files randomly. This makes a traditional random fuzzer often ineffective in finding bugs that hide deep in the execution and get stuck during fuzzing.
Existing fuzzers like AFL~\cite{AFL} and its many variants~\cite{fioraldi2020afl++,gan2018collafl,bohme2017directed,peng2018t, 8233151, rawat2017vuzzer, li2017steelix, chen2018angora, libFuzzer, manes2020ankou, pham2020aflnet, yue2020ecofuzz}use evolutionary algorithms (and sometimes alternative optimization algorithms~\cite{8233151,cha2015program,lyu2019mopt}) to generate tests inputs. The optimization goal is to maximize the code coverage of the program so as to maximally reveal potential security vulnerabilities. In particular, AFL-based fuzzers instrument the program under test to monitor the coverage of each program execution, record the test inputs that cover different branches, select the promising test inputs (called seeds) and mutate the selected seeds in the hope of improving the coverage. The process repeats until a time budget is exhausted. This simple strategy often allows us to efficiently cover a large number of code blocks in the program. \emph{Its effectiveness, however, often deteriorates over time}. 

There are two main reasons. 
%it only provides a coarse-grained observation of program execution and have two main drawbacks which makes AFL can only discover deep execution paths by sheer luck. 
First, existing AFL-based fuzzers select test inputs which cover new branches as seeds. While such a strategy is effective initially, overtime test inputs which cover new branches become few and far in between, which renders such a strategy ineffective.  
To solve this problem, \emph{we need a systematic and adaptive way of identifying the most ``rewarding'' (in terms of covering those un-covered branches) test inputs as seeds}. Second, after certain seeds are selected, existing AFL-based fuzzers apply a rich set of mutation operators on the seeds to generate new inputs. Similarly, such a strategy becomes ineffective overtime. After covering the easy-to-cover branches, covering the remaining branches often require specific mutation on specific ``bytes'' in the inputs. To address this problem, \emph{we need a way of knowing where and how to apply the mutation operators in order to cover those uncovered branches}. %Some AFL variants aim to overcome this problem by XXX \todo{review existing attempts on this problem and say why they are not good - probably because they don't perform a global analysis}. 

%Some AFL variants aim to overcome this problem by XXX

%cannot effectively select the most promising inputs to mutate from the discovered paths. In other words, the fuzzers do not know that selecting which torrent file will bring greater potential benefits. Secondly, after selecting the seed file, for different mutators, the fuzzers don't know which position the mutation can make the most efficient test. Fortunately, we can increase the efficiency of AFL-like fuzzers manifold by accounting for information that answers the questions above.

There are multiple attempts on addressing these two problems in the literature. 
% \todo{Re-organize the related work here in terms of how they address problem 1 and 2 above}. 
To select seeds that are likely to cover new branches, LibFuzzer~\cite{libFuzzer}'s heuristic is to select newly generated tests. AFLfast\\~\cite{8233151} prioritizes the seeds that can trigger the less-frequently visited pathes. Entropic~\cite{bohme2020boosting} selects seeds that carry more program information according to certain entropy measure. Fairfuzz~\cite{lemieux2018fairfuzz} locates and selects the seeds which trigger low-probability edges. Cerebro~\cite{li2019cerebro} selects seeds based on multiple factors such as code complexity, execution time and coverage. While being effective to some extent, these approaches lack a global view of the program under fuzzing and thus often miss the most `rewarding' seeds over time. For instance, a low-probability edge may not be as ``rewarding'' as a high-probability edge if the latter leads to a large number of uncovered branches. 

% \todo{review existing attempts on this problem and say why they are not good - probably because they don't perform a global analysis}. 
To selectively apply mutations, Steelix~\cite{li2017steelix}, REDQUEEN~\cite{aschermann2019redqueen} and other works~\cite{cadar2008klee, stephens2016driller, king1976symbolic, sen2005cute, molnar2009dynamic, godefroid2008automated, cha2015program, cadar2013symbolic, cadar2011symbolic, godefroid2005dart, khurshid2003generalized, choi2019grey} proposed to perform dynamic taint analysis to determine the specific bytes in the input for solving the so-called ``magic bytes'' problem.
MOpt~\cite{lyu2019mopt} utilizes a customized particle swarm optimization to guide mutations scheduling. 
Besides, machine learning has been introduced to improve the performance of fuzzing in recent years~\cite{chen2018angora, wang2010taintscope, haller2013dowsing, neugschwandtner2015borg, rawat2017vuzzer, li2017steelix, harman2009theoretical}. For instance, RNN fuzzer~\cite{rajpal2017not} uses a recurrent neural networks (RNNs) to predict whether an input can reach the target program block to filter out uninteresting test cases. Based on a fairly similar idea, FuzzGuard~\cite{zong2020fuzzguard} achieves good results in directed fuzzing. Neuzz~\cite{she2019neuzz} uses neural networks to smooth the target program and guides the input variation through gradients. However, these approaches only address the problem partially, i.e., only to identify the relevant bytes but not how to select the mutation operator (among the many choices). 
% \todo{This sentence is not convincing - since we too have the overhead. Shall we say: However, their approach only addresses the problem partially, i.e., they are only able to identify the relevant bytes but not how to mutate them.} 

For hybrid fuzzers, Driller~\cite{stephens2016driller} uses concolic execution to explore new paths when it gets stuck on superficial ones.
QSYM~\cite{yun2018qsym} using dynamic binary translation to efficiently solve symbolic emulation.
% \todo{Similarly review other related works on how they address problem 2 and say why they are not ideal.} 

% \todo{I don't think we need to review those hybrid fuzzing approaches here. We compare with them in the related section only. }
%Unfortunately, these machine learning-based methods lack the overall analysis of the program and can not select the appropriate seed file according to the fuzzing. 
% Existing methods often cause the fuzzer to be distracted during the testing, that is, the selected seed file and mutation strategy result in most of the generated input scattering on the path that has been fully fuzzed, and only a small amount of generated input can reach those basic blocks that can potentially bring more gains. 

In this work, we introduce \tool, a novel framework to address the two problems systematically. First, in order to draw the fuzzer's attention to those test inputs that are most rewarding, \tool quantifies the `reward' of covering each basic block in the program through a lightweight global analysis. Intuitively speaking, \tool estimates the reward based on the probability of covering uncovered branches and the number of them. Second, in order to draw the fuzzer's attention to specific mutations on specific bytes of the seeds that are most rewarding, \tool trains a model which predicts whether a certain ``rewarding'' block will be covered given certain mutation on certain bytes of a seed. This is achieved by training an explainable deep learning model with attention mechanism which is learned and updated periodically.

%Next, based on the quantitative reward, \tool screens out the most noteworthy blocks, and introduces a deep learning model with attention mechanism. While classifying whether the input can reach the target node, it extracts the impact of different mutators and parameters on the execution of the program to further guide mutation.

\tool has been implemented on top of AFL. We systematically evaluate \tool with multiple experiments, comparing \tool with 4 related state-of-the-art fuzzers on 13 real-world programs. The results show that \tool significantly outperforms all existing fuzzers both at achieving higher edge coverage and finding new bugs. In particular, \tool achieved 2X edge coverage and 4X bugs detected than AFL over 24 hours' run. More importantly, thanks to the seed selection strategy and the attention-based deep learning model, \tool consistently improves its coverage over time (whereas existing fuzzer's effectiveness drops significantly after 24 hours), i.e., achieving 50\% more coverage than AFL after 5 days of fuzzing. 
% \todo{Add detailed statistics if there is any.}

%Thanks to our proposed method of focusing the attention of the fuzzer, the \tool can run efficiently during the fuzzing process and focus on breaking through the bottleneck. We tested the \tool in the mainstream LAVA-M dataset and real-world programs(libjpeg, mupdf, libxml, etc.) that require various types of input. The result shows that \tool consistently outperforms all the other fuzzers by a wide margin both in terms of detected bugs and achieved edge coverage.

In a nutshell, our technical contributions are as follows.
\begin{itemize}
    \item We propose a lightweight global analysis to dynamically and adaptively identify the most ``rewarding'' test inputs as seeds during the fuzzing process.
    \item We propose to use explainable deep learning models with attention mechanisms learned from the massive fuzzing data to identify effective mutations on specific bytes of the identified seeds.
    \item We design, implement, and evaluate \tool and demonstrate that it significantly outperforms 4 state-of-the-art fuzzers on a wide range of real-world programs.
\end{itemize}

% The remainder of the paper is organized as follows. We review relevant background in Section~\ref{sec:back} and present a motivating example in Section \ref{sec:mov}. Then, the framework of \tool is presented in Section \ref{sec:fr} with implementation details provided in Section \ref{sec:id}. We then evaluate \tool in Section \ref{sec:ev}. Lastly, we present related works in Section \ref{sec:re} and conclude in Section \ref{sec:con}. 
% % \todo{Add details.}

\section{Background}
\label{sec:back}
% In this section, we review background and define our problem. 

\subsection{Coverage-guided Fuzzing}

We start with formalizing the grey-box fuzzing problem.

\begin{definition}
A program is a labeled transition system $\mathcal{P} = (B, init, V, \phi, GC, T)$ where
\begin{itemize}
    \item $B$ is a finite set of control locations\footnote{We use `control location' and `basic block' interchangeably throughout.};
    \item $init \in B$ is a unique entry point of the program;
    \item $V$ is a finite set of variables;
    %\item $\phi$ is a predicate capturing the set of initial valuations of $V$;
    \item $GC$ is a set of guarded commands of the form $[g]f$, where $g$ is a guard condition and $f$ is a function updating valuation of variables $V$. $f$ represents a basic code block in general.
    \item $T: B \times GC \rightarrow B$ is a transition function. 
\end{itemize}
\end{definition}
Note that for the sake of presentation, the above definition assumes a flattened program structure without functions, classes and packages. We leave the details on how function calls are handled in the implementation section. 

A concrete execution (a.k.a.~a test) of $\mathcal{P}$ is a sequence
\[\pi = \langle (v_0, b_0), gc_0, (v_1, b_1), gc_1, \cdots, (v_k, b_k), gc_k, \cdots \rangle,\] 
where $v_i$ is a valuation of $V$, $b_i \in B$, $gc_i = [g_i]f_i$ is a guarded command such that $(b_i, gc_i, b_{i+1}) \in T$, $v_i \vDash g_i$, and $v_{i+1} = f_i(v_i)$ for all $i$, and $v_0 \vDash \phi$ and $b_0 = init$. We use $\Pi$ to denote a set of tests. We say a test $\pi\in\Pi$ covers a control location $b$ if and only if $b$ is in the sequence. A control location $b$ is reachable by $\Pi$ if and only if there exists a concrete execution $\pi\in\Pi$ which covers $b$. 
% The initial variable valuation $v_0$ is also referred to as a seed file.
\begin{definition}
\textbf{Pre-dominant Blocks} Given a basic block $b \in B$, 
we define the set of $b$'s pre-dominant blocks $D_b=\{b'|\ b'\in B \ \& \ \exists \ g:(b', gb', b)\in T\}$. 
%$b'$ is a pre-dominant block of $b$ if $b'$ can reach b with one step in the program's DTMC. 
\end{definition}

Intuitively, the set of pre-dominant blocks $D_{b}$ are those blocks which could transit to $b$ in one step. 
%\begin{algorithm}[t]
%	\caption{AFL-style Fuzzing Algorithm}
%	\label{afl_alg}
%	Let $\mathcal{P}$ be the target program\;
%    Let $\mathcal{B}$ be the set of bugs\;
%    
%    \For{$seed\in seed\_pool$}{
%        $length$ = $len(seed)$\;
%        $seed\_coverage$ = $Execute(\mathcal{P}, seed)$\;
%        \For{$iterations \leq limit$} {
%            \For{$mutator \in mutator\_list$} {
%                \For{$loc \leq length$} {
%                    $input$ = $mutate(seed)$\;
%                    $cov, result$ = $Excute(\mathcal{P}, input)$\;
%                    \If{$result == Crash$} {
%                        $\mathcal{B}.append(input)$
%                    }
%                    \If{$HasNewCov(cov)$} {
%                        $seed\_pool.append(input)$
%                    }
%                }
%            }
%        }
%    }
%    \Return $\mathcal{B}$\;
%\end{algorithm}

\begin{figure}[t]
  \centering
  \includegraphics[width=.9\linewidth]{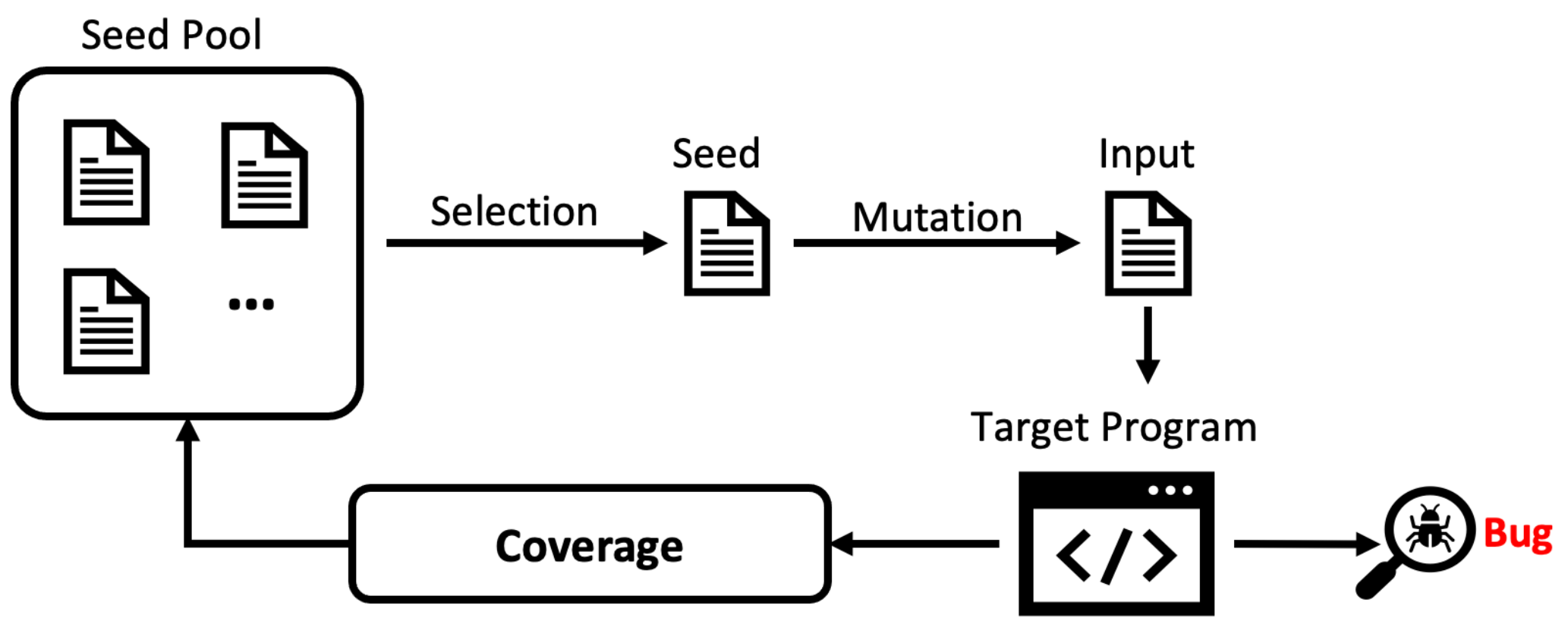}
  \caption{Coverage-guided Fuzzing Overview.}
  \label{covguided}
\end{figure}

Grey-box fuzzers like AFL~\cite{AFL} are designed to generate a set of test inputs $\Pi$ which covers as many edges of the target program as possible with the hope of triggering bugs. We summarize the overall process of such fuzzers in Figure~\ref{covguided}. First, the target program is instrumented to obtain program coverage information during the fuzzing process. Second, to maximize reachable $B_b$, the fuzzer selects and mutates a test from a seed pool such that the mutated input would incur a different concrete execution of the program (covering new control locations). Afterwards, the program coverage information is updated and those tests incurring a different execution are prioritized and added into the seed pool. Afterwards, a new iteration starts.

% In terms of input generation, AFL-style fuzzers mostly use seed file pools and random mutations to generate diverse inputs. 

%In general, a seed input $s$ can be regarded as a sequence of bytes $s = \langle byte_{0}, byte_{1},\cdots,byte_{N} \rangle$ derived from the valuation of $V$. Given $s$, commonly used mutation operations are summarized as follows.

% \todo{The formalization below is whatever problematic. For instance, the result is not always a test; some of the operators are not defined; and so on. I would suggest to simply explain using texts.}

% \begin{itemize}
% 	\item $Bitflip(s, k) = s.bit_{k} \oplus $ 0x1, where $bit_{k}$ is the $k$-th bit of $s$.
% 	\item $Arth_{\pm}(s, k, x) = s.byte_{k} \pm x$
% 	\item $Replace(s, k, x) = \langle byte_{0},\cdots,byte_{k-1},x,byte_{k+1},\cdots,byte_{N}\rangle$
% 	\item $Insert(s, k, s') = \langle byte_{0},\cdots,byte_{k},s',\cdots,byte_{N+k}\rangle$, where $s'$ is another sequence of input.
% 	\item $Delete(s,k,l) = \langle byte_{0},\cdots,byte_{k},byte_{k+l},\cdots,byte_{N-1}\rangle$
% \end{itemize}

The challenging problem to be solved by fuzzing is to identify the most `rewarding' seeds and mutations so that program edges are covered efficiently. The problem is highly non-trivial due to the large search space, i.e., there are often many test cases which could serve as seeds and, given a seed, there are many possible mutations\footnote{We refer the reader to~\cite{AFL} for details due to space limit.} (transform, deletion, and splicing, etc) as well. For instance, given a test seed of $N$ bytes, the number of mutations defined by the AFL mutation operators is $29^{N}$. 
% We remark that there are extensive works on how to select the seeds and mutations~\cite{rawat2017vuzzer, li2017steelix, 8233151, stephens2016driller, zhou2020zeror, peng2018t, chen2018angora}. These existing approaches are based on intuitive heuristics without a global analysis on how rewarding a selected seed and the mutation would be. 
% \jy{add the most related references here} 

% Coverage-guided fuzzers like AFL use the input generated from these mutations to run the program and observe these conditions at the same time: 1. Whether this execution triggers a new program branch 2. Whether it triggers a bug. If the current execution triggers a new program branch, the fuzzer will take this input as a new seed and add it to the seed pool. If this execution triggers a bug, the current input will be added to the bug set, and the loop will go back and forth. 

% In the entire fuzzing process, although AFL-style fuzzers have a certain insight of the program coverage, they are completely program-agnostic in terms of input generation. 

% Specifically, they have no idea where the seed should be mutated and which mutator should be used. Whether there is a new input gain can only depend on the previously obtained seed.

\subsection{Deep Learning and Attention Mechanism}

%Deep learning is a class of machine learning algorithms that uses multiple layers to progressively extract higher-level features from the raw data. The information obtained in the learning process is of great help to the interpretation of data such as text and images. Its ultimate goal is to allow the machine to have the ability to analyze and learn, and be able to recognize and process data. The adjective "deep" in deep learning refers to the use of multiple layers in the network. By designing and establishing an appropriate number of neuron computing nodes and a multi-layer computing hierarchy, deep learning can establish a functional relationship from input to output. In the fuzzing scenario, deep learning can establish the relationship between input distribution and program coverage through massive fuzzing data. The upper part of Figure~\label{attoverview} depicts a simple deep learning example, which first extracts three-dimensional features from the input picture through embedding and feature extraction layers(e.g, convolutional layers) and then classifies the results through classification layers(e.g,fully connect layers).

Deep learning is a class of machine learning algorithms that use multi-layer neural networks to abstract high-level features from the input data. Different from traditional machine learning algorithms, deep learning models automatically learn features from data rather than using handcrafted features. 
%\todo{Figure is missing}
%\textcolor{red}{Take image classification problem as an example, 
%The upper part of Figure~\ref{attoverview} depicts a simple deep learning example. 
%classic method firstly represents an input figure using its RGB matrices, and a feature extraction layer with convolutional neural network structure is applied to extract its abstract features.} The extracted features are then fed into a final classification layer to support the final decision~\cite{krizhevsky2012imagenet}. 
This end-to-end deep learning framework does not require complex manual feature engineering and has shown distinct advantages in various areas, including computer vision~\cite{krizhevsky2012imagenet}, natural language processing~\cite{collobert2011natural} and progressively applied in computer security~\cite{firdausi2010analysis}.

%In this work, we will further utilize attention model to capture the contributions of each component with attention weights, which can improve the fuzzing performance and provide the interpretability of our deep learning model.

%In this work, we will utilize attention model (inspired by human's visual perception mechanism) to help us understand the contribution of each component in fuzzing and use that to boost the fuzzing process. 

% \begin{figure}[t]
%   \centering
%   \includegraphics[width=\linewidth]{attoverview.pdf}
%   \caption{Deep learning and attention model overview}
% \label{attoverview}
% \end{figure}

% \jy{Jie can help on this part}
 
Further, the attention mechanism enables a neural network model to distinguish the contributions of different input segments in the model decision process~\cite{bahdanau2014neural}, by explicitly assigning a unique weight for each basic unit of input data, and calculating the representation of the input data using weighted sum of all basic units.
% Figure~\ref{model} demonstrates the attention mechanism in a deep learning framework. In a deep neural network with attention mechanism, an attention layer follows the feature extraction layer to calculate the weight of each basic unit in the input. The weight distribution of the input is called the attention mask, which represents the importance of different input parts. The final representation of the input is the weighted sum of the input basic units.
Deep learning with attention mechanism is particularly relevant to the above-defined fuzzing problem for the following reasons. To identity the most rewarding mutation given a seed, we need to efficiently predict what mutations are more likely to cover certain uncovered program edges. Given the complex relation between the test inputs and program coverage, a powerful model like a deep learning model is necessary. More importantly, the attention mechanism allows us to `understand' the coverage of particular `magic' bytes, which provides effective guidance on future mutations.
% \vspace{1mm}
\subsection{Problem Definition}
% \sj{We need a subsection on problem definition here. 
% The problem is to find an effective way of selecting/mutating seeds such that the coverage is improved. }
% For fuzz testing, how should we focus the fuzzer's attention to improve efficiency?
% First of all, a large number of seed files will be generated during the fuzzing process, and different seed files can cover different program branches. The problem is in the seed file pool, which seed files should we focus on? In other words, how should we select seed files in the seed file pool to perform mutation?
% Second, the mutation strategy is quite complicated. Fuzzer can perform different kinds of mutations in different positions of the seed file to generate input. The problem is which mutator and seed file positions should we focus the fuzzer's attention on? In other words, how to choose a suitable mutator and mutate in a suitable seed file position to improve the efficiency of fuzzing?
% \begin{definition}
% Let P be a program under test. Let $\Mu$ be a set of mutation operators. The grey-box fuzzing problem is to develop a strategy which is composed a function for seed section and a function of mutation selection such that the coverage is maximized.
% \end{definition}
We summarize our core problem as how to \emph{pay better attentions} whilst fuzzing to improve program coverage effectively and consistently from the following main aspects:
\begin{itemize}
    \item How to pay attention to the most rewarding seeds?
    \item How to pay attention to the most rewarding bytes and mutations?
\end{itemize}

\section{\tool FRAMEWORK}
\label{sec:fr}
% In this work, we take advantage of the power of deep learning (from massive fuzzing data) to improve
% program coverage efficiently. Such a data-driven approach has two
% major advantages compared to traditional techniques like dynamic
% stain analysis and symbolic execution. 

% Firstly, deep learning can help the fuzzing engine to capture the relationship between program execution and input distribution, which could further guides the mutation strategy to improve program coverage. Secondly, the power of deep learning is dynamically increasing as we obtain more fuzzing data, which in turn speeds up the fuzzing process. In addition, the attention mechanism can further improve the deep learning's performance and enhance its interpratability as well.

%\begin{figure}[h]
%  \centering
%  \includegraphics[width=\linewidth]{Overview.pdf}
%  \caption{Overview of XXX workflow}
%  \label{overview}
%\end{figure}
\begin{figure*}[t]
  \centering
  \includegraphics[width=.73\linewidth]{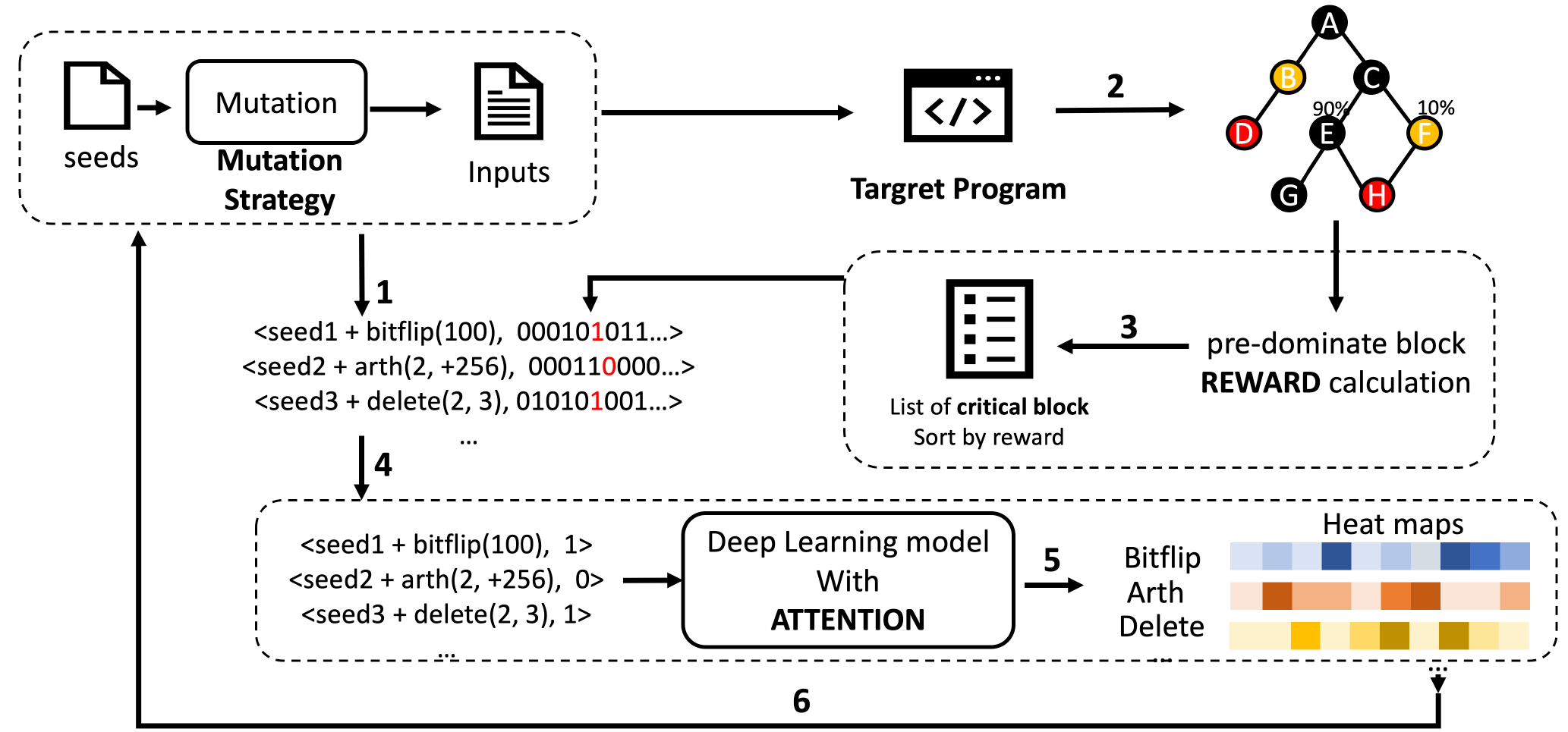}
  \caption{\tool overview.}
  \label{overview}
  \Description{overview}
\end{figure*}

In this section, we present details of \tool. An overview of \tool is shown in Figure~\ref{overview}. \tool includes four main phases: data collection, reward calculation, model training and mutation strategy updating. In the data collection phase, \tool adopts a carrier fuzzer to generate inputs and records the seed files and mutations used. \tool also tracks the coverage achieved by each test (step 1 and step 2 in Figure~\ref{overview}). Over time, the fuzzing process often gets stuck and the coverage is difficult to improve. 
% However, the effort of the carrier fuzzer is scattered throughout the program, not focusing on breaking the bottleneck. 
% In this work, 
Then, \tool is activated.
%helps the fuzzer pay better attention on the bottleneck blocks as follows. 
The core idea is to adopt deep learning with attention to predict whether a seed and mutation combination can cover certain basic block based on a basic block's coverage data.
% and obtain the corresponding heat maps. 
However, due to the large number of uncovered basic blocks, it is costly to train a model for each of them. Apart from that, we are not able to learn an effective model since we do not have any positive labeled data for the uncovered blocks. To address the challenges, as shown in Figure~\ref{overview},in step 2, we build an abstraction of the program in the form of a (labeled) discrete time Markov Chain (DTMC). Then, step 3 aims to find out \emph{critical blocks} based on the DTMC and step 4 prepares the respective fuzzing data.
%\tool calculates a reward for each uncovered block based on the probability estimated from the fuzzing records and selects a set of blocks as the targets. Then, \tool uses the control flow graph (CFG) to find the pre-dominant blocks of the target blocks as \emph{critical blocks} (step 3) and prepare the respective fuzzing data (step 4) for the following model training phase. The key intuition is that covering these \emph{critical blocks} will significantly increase the probability of reaching the target uncovered blocks. 
% According to the size of the reward, we select critical blocks for the next stage of training (step 3 and step 4 in figure~\ref{overview})
%  and . In the model training phase, for each high reward basic block, 
Next, step 5 aims to get the heat maps of the seed file under different mutators to provide guidance on selecting the more valuable bytes and corresponding mutators by training an attention model.
%In particular, \tool trains a deep learning model with attention mechanism to obtain the heat maps of the seed file under different mutators (step 5) to provide guidance on selecting the more valuable bytes and corresponding mutators.
% In the mutation strategy updating phrase, using the obtained heat map, \tool updates the fuzzer strategy by skipping mutations on high heat positions. 
%Finally, the updating phrase is performed when fuzzing breaks through the current bottleneck and encounters new ones. The above steps are iterated to update the overall fuzzing strategy (step 6 in figure~\ref{overview}).
% }
The above process continues until the current bottleneck is overcome. 

% Next, we present the details of each phase.
% \jy{this paragraph is clear, the writing could be improved though. besides, for concepts like pre-dominating nodes, reward etc, we need to give formal definitions later}

\subsection{Data Collection}
 
% we mainly collect the input used for each execution and the corresponding program coverage information. AFL-like fuzzer often generates rich input through seed file and a series of mutations.

In the data collection stage, \tool collects relevant information on test inputs generated by the carrier fuzzer, i.e., the seed file and mutations used. For coverage, \tool records the AFL bitmap to track which basic blocks are triggered.
%\tool records the input generated by the carrier fuzzer together with the seed file, the mutator used, and the mutation parameters. For coverage, \tool records the AFL bitmap to track which basic blocks are triggered. In summary, for each program execution, \tool collects a set of data in the form of a quadruple $(seed\ file, mutator, mutation\ parameters, bitmap)$.

\begin{example}
For the program in Figure~\ref{abstraction}, AFL performs an $Arth_{+}$ operation on the seed file $\langle 0, 5 \rangle$ with parameter 5 on the first byte, and gets the final input $\langle 5, 5 \rangle$. This input can cover blocks 1, 2 and 3 in the program. So we record ($\langle 0, 5 \rangle$, arth+, 5; 111000). 
% Similarly, after a series of mutations, we can get ($\langle -5, 0 \rangle$, arth+, 5; 110101), ($\langle 0, 0 \rangle$, bitflip, 5; 100101), ($\langle 0, 0 \rangle$, interest, 0xFF; 100101) etc. 
After a series of mutation operations, we can get the same form of data, such as ($\langle -5, 0 \rangle$, arth+, 5; 110101), ($\langle 0, 0 \rangle$, bitflip, 5; 100101), ($\langle 0, 0 \rangle$, dictionary, 0xFF; 100101).
\end{example} 
% For example, 

\subsection{Reward Calculation}

As mentioned above, to effectively solve the fuzzing problem, we need a systematic way of identifying the most rewarding seeds and mutations. Intuitively, a test case is most rewarding if it leads a maximal improvement of the edge coverage. In the following, we present a lightweight approach which allows us to systematically compute the reward of covering a basic block. Note that the reward is then used as a guide to select seeds and mutation, i.e., those which are predicted to cover the basic blocks with highest rewards. Our approach is inspired by ~\cite{wang2018towards,wang2017improving}, which enables us to build a discrete-time Markov Chain (DTMC) abstraction of the program from the collected fuzzing data. Specifically, 
\begin{definition}
A (labeled) discrete-time Markov Chain (DTMC) is a tuple $\mathcal{M} = (B, Pr, \mu)$ where $B$ is the set of basic blocks in $\mathcal{P}$; $Pr: B \times B \rightarrow \mathbb{R^+}$ is a labeled transition probability function such that $\Sigma_{b' \in B} Pr(b, b') = 1$ for all $b \in B$; and $\mu$ is the initial probability distribution such that $\Sigma_{b \in B}\mu(b) = 1$. 
\end{definition}

% above, after a period of fuzzing, the program's coverage will often reach a bottleneck, and it is difficult for the fuzzer to generate new inputs to obtain coverage improvements efficiently. To help the fuzzer choose a suitable mutation strategy, \tool introduced a deep learning method for the blocks that have not been covered. 
% However, the number of uncovered blocks is enormous, and the overhead caused by learning each block is unacceptable. Therefore, we try to filter out the critical blocks from a large number of uncovered blocks. 

% On the basis of defining the program as a labelled transition system, we try to transform the program into a (labeled) discrete time Markov Chain (DTMC), which is defined as follows:

\begin{figure*}[thb]
\begin{minipage}{.8\textwidth}
\centering
\includegraphics[scale=.75]{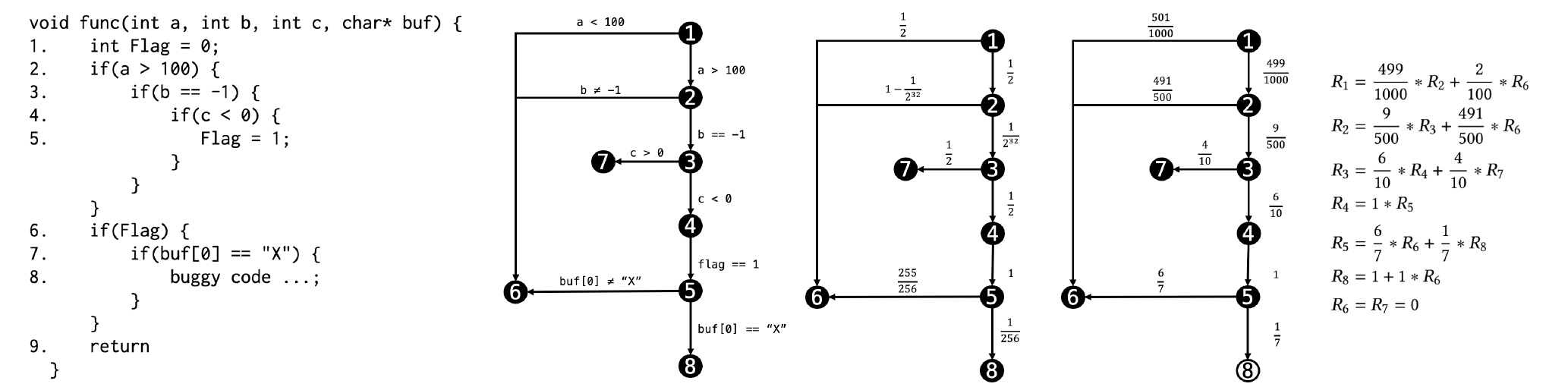}
\end{minipage}

% \begin{minipage}{.45\textwidth}
% {\small \begin{align*}
% R_1 & =  \frac{499}{1000}*R_2 + \frac{2}{100}*R_6 \\
% R_2 & =  \frac{9}{500}*R_3 + \frac{491}{500}*R_6 \\
% R_3 & =  \frac{6}{10}*R_4 + \frac{4}{10}*R_7 \\
% R_4 & =  1*R_5 \\
% R_5 & =  \frac{6}{7}*R_6 + \frac{1}{7}*R_8 \\
% R_8 & =  1 + 1*R_6 \\
% R_6 & =  R_7 = 0\\
% \end{align*}}
% \end{minipage}
\caption{Program abstraction}
\label{abstraction}
\end{figure*}

Naturally, we can abstract a program into a DTMC if we impose an initial distribution on its initial states, where each control location in the program becomes a state in the DTMC, and each edge between two control locations are associated with a conditional probability. For example, a program shown on the left side of Figure~\ref{abstraction} can then be transformed into the DTMC on the right~\cite{wang2018towards}. The key to construct the DTMC is to estimate the conditional probabilities between edges from the fuzzing records as follows:

% We use the concept of reward to measure the importance of a basic block. The higher the reward value, it means that covering this block can bring us higher benefits, In other words, it is easier to cover more blocks.
% To calculate the reward of the node, we first abstract the program into a (labeled) discrete-time Markov Chain (DTMC). We write $Pr(b, b')$ to denote the conditional probability of visiting $b'$ given the current state $b$. The conditional probability $Pr(b, b')$ is also called as one-step transition probability.

%However, it is obvious an ideal situation to obtain each edge's transition probability accurately. It is difficult for us to know each transition's probability accurately in actual scenarios, or doing so will cause unacceptable overhead. The simplest solution is: for basic block B, set the probability of each edge starting from it to $\frac{1}{n}$, where n is the total number of block b can reach with one step. Obviously, there are considerable errors in this method. For example, the probability of block 1 to block 2 in the example is $\frac{1}{2^{32}}$, and $\frac{1}{2}$ is $2^{31}$ times larger. To solve this problem, we continuously estimate the actual probability to reach the optimum by counting the coverage at runtime. With the execution of the program, the edge probability will gradually approach the true probability. 

\begin{definition}
Let $\#(b1, b2)$ denote the total number of times $b1$ transits to $b2$ in the fuzzing process and $\#b1$ represents the total number of executions of $b1$, $n$ is the total number of outgoing edges of $b_1$ in the CFG. Then the conditional probability from $b_1$ to $b_2$ is then $Pr(b1, b2)$ = $\frac{1 + \#(b1, b2)}{\#b1 + n}$. 
\end{definition}

% Intuitively, if the probability of a state being executed is small and the number of its child nodes is large, then covering this node is more likely to bring us higher benefits. With reference to ~\cite{optimal}, our definition of reward is as follows:
The reward of a basic block is defined as follows.

\begin{definition}
Let $R_b$ where $b \in B$ be the reward of visiting b.  $t \in T$ be the block set that b can reach with one step. We build an equation system as follows. 
\[R_b=\begin{cases}
               1 + \Sigma_{t \in B} \{Pr(b, t) \times R_t\} & \text{if $b \not \in visited$} \\
               \Sigma_{t \in B} \{Pr(b, t) \times R_t\} & \text{if $b \in visited$}
            \end{cases}	
\]
\end{definition}

With an estimated DTMC and the above equation system, we can calculate each basic block's reward with the help of the program's CFG.
Note that for those indirect calls that are unable to be extracted by static analysis, we use the dynamic fuzzing data to complement the CFG of static analysis.
We omit the details of solving the equation system (which is efficient, e.g., in seconds) and refer interested readers to~\cite{wang2018towards} fo details. 

\begin{example}
We use the program in Figure~\ref{abstraction} as an example to illustrate the reward calculation process. As shown in Figure~\ref{abstraction}, after 998 program executions, we can estimate the DTMC from the basic blocks coverage information. The equation system can be built on the bottom of Figure~\ref{abstraction}. By solving the equation system, we have the rewards of covering each basic block as $R_1=0.001$, $R_2 = 0.002$, $R_3 = 0.086$, $R_4 = 1.333$, $R_5 = 0.143$, $R_6 = 0$, $R_7 = 0$ and $R_8 = 1$.

% We dynamically define the probability of a basic block $P_{b}$ as $\frac{\#b}{\#Total}$, where $\#b$ represents the number of times $b$ are executed in the fuzzing process and $\#Total$ represents the total number of fuzzing execution.
%In order to further filter the data obtained by reward for use in the deep learning stage, we define the pre-dominant block as follows.
\end{example}

% \begin{figure}[t]
% \begin{minipage}{.45\textwidth}
% \centering
% \includegraphics[width=.9\textwidth]{estimation.pdf}
% \end{minipage}

% \begin{minipage}{.1\textwidth}
% {\small \begin{align*}
% R_1 & =  \frac{499}{1000}*R_2 + \frac{2}{100}*R_6 \\
% R_2 & =  \frac{9}{500}*R_3 + \frac{491}{500}*R_6 \\
% R_3 & =  \frac{6}{10}*R_4 + \frac{4}{10}*R_7 \\
% R_4 & =  1*R_5 \\
% R_5 & =  \frac{6}{7}*R_6 + \frac{1}{7}*R_8 \\
% R_8 & =  1 + 1*R_6 \\
% R_6 & =  R_7 = 0\\
% \end{align*}}
% \end{minipage}
% \caption{Reward Calculation Example}
% \label{estimation}
% \end{figure} 

Once we know the reward of covering each uncovered basic block, \tool then selects the top $k$ percent of them as target uncovered blocks $B_c$ (most rewarding).
% $B_c$ to pay more attention to by deep learning. 
As mentioned, since we have no positive labeled data for the uncovered blocks, we obtain the pre-dominant blocks of each target block as $pre(B_c)$ by static analysis.
The blocks in $pre(B_c)$ are our target for fuzzing. To further reduce the number of targets (and thus reduce the number of mutants to be generated), we filter those blocks in $pre(B_c)$ which have a high probability of being reached according to the DTMC (since they hardly need much assistance).
Note that among $pre(B_c)$, we prefer those which has a low probability to reach as more interesting \emph{critical blocks}. In practice, we omit those pre-dominant blocks which has a probability higher than a threshold $k'$. We use $B_{critical}$ to denote the finally selected pre-dominant blocks for deep learning.  

Overall, with the help of the reward mechanism, the goal becomes to generate test cases which is likely to cover $B_{critical}$. 
Now we know which block to cover, we further need a way of predicting which seed and mutation would cover that block.

\subsection{Training Attention Models}
\label{tam}

The above two phases enable \tool to pay attention to the most valuable basic blocks ($B_{critical}$). Next, we build a deep learning model to systematically predict if a seed-mutation combination is likely to cover a certain code block.
%perform model training on these blocks to better improve coverage. 
We choose an attention based deep learning model which can extract features from inputs automatically and making the correct classification (of whether a block will be covered). More importantly, the attention mechanism enables our model to distinguish the impact of different mutators and parameters on each byte for different seed files. In particular, we prepare the training data which are composed of two parts: one is the seed file, and the other is the mutator and the corresponding mutation parameters. 

Specifically, we convert the seed file into a series of vectors $\{ \mathcal{X}_{D\times N}=\langle v_1,v_2,\cdots,v_{N} \rangle,\ v_{m_{D\times 1}},\ v_{p_{D\times 1}} \}$, where $D$ is the dimension of bytes embeddings or pixel vector, and $N$ is the largest size of the collected seed files. Note that for those inputs whose size is less than the maximum length, we padding the input to length $N$. Next, we adopt customized models for different kinds of programs to extract the relevant features depending on their inputs. For instance, for programs that take image as input, traditional convolutional neural networks (CNNs)
are used. 
% Refer to image recognition, features of an object in an image are represented by combinations of several pixels (i.e., elements in the input vector), which could be well extracted by the CNN models. 
For programs that take byte sequences as input (such as inputs which are XML files or JSON string), recurrent neural networks (RNNs) are used to capture the sequential information. These two networks have been widely used in image feature and byte sequence feature extraction. After feature extraction, we can get the vector of extracted features as $\mathcal{U}_{D'\times N} = \langle v'_1,v'_2,\cdots,v'_{N}\rangle = f_{conv}(\mathcal{X}_{D \times N})$.

\begin{figure}[t]
\centering
\includegraphics[width=\linewidth]{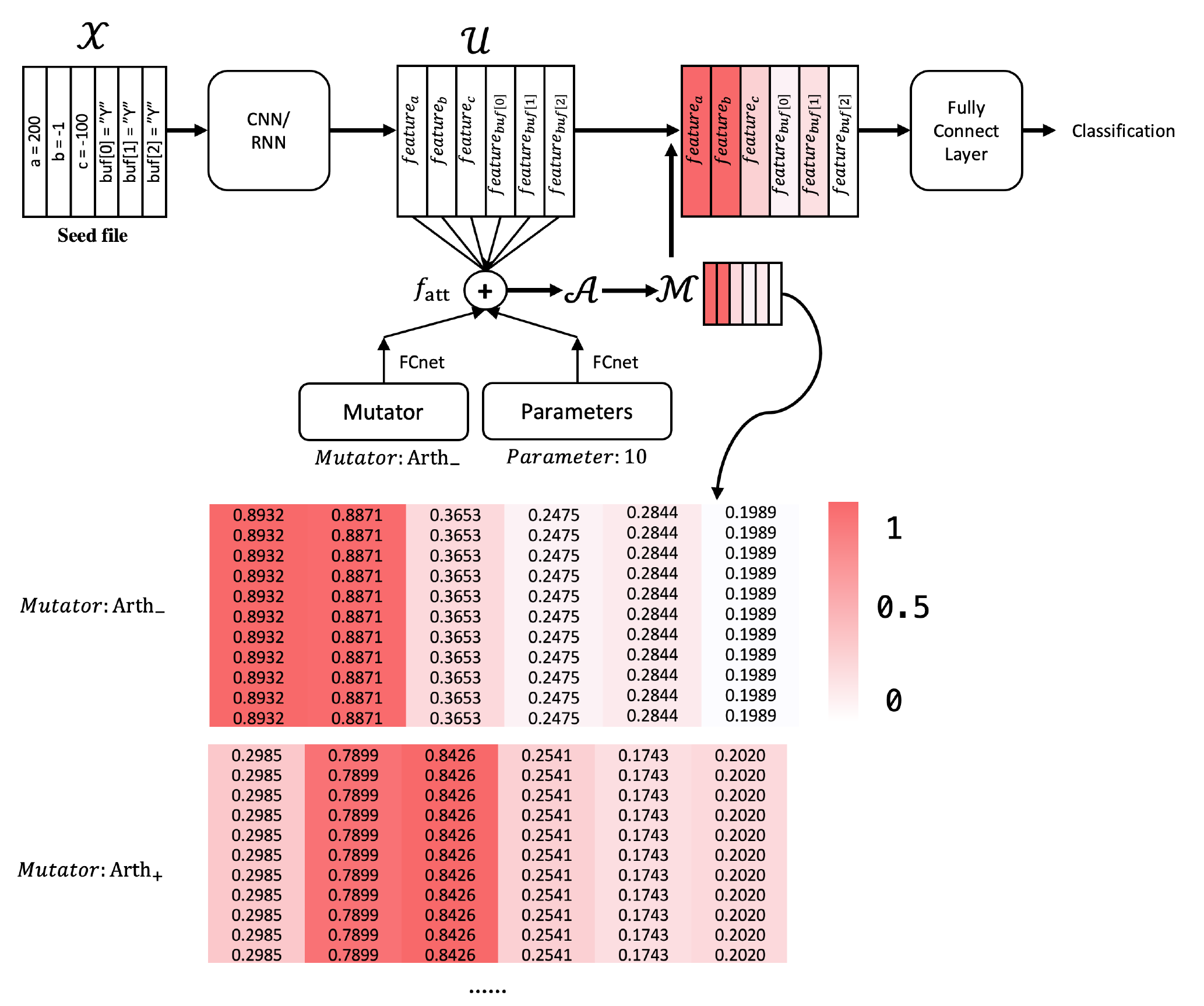}
\caption{Attention model}
\label{model}
\end{figure}

% Whether through CNN or RNN, simply extracting features from the seed file for classification is not sufficient. Although thousands of inputs are generated during the fuzzing process, seed files are often limited. 
% To increase the size of training data, we add an additional operator <mutator, parameters> to generate more training cases. 
To further utilize the information of mutations, we add in the mutation operator <mutator, parameters>. Then, as shown in Figure~\ref{model}, for each vector in $\mathcal{U}$, we calculate an attention weight $\alpha$:

%\begin{figure}[t]
%  \centering
%  \includegraphics[width=\linewidth]{heatmap.pdf}
%  \caption{Example of heat map.}
%  \label{heatmap}
%\end{figure}

\begin{equation}
\alpha_i=f_{att}(\mathcal{U}_{D'\times N},v_{m_{D \times1}},v_{p_{D\times1}})
\end{equation}

The function $f_{att}$ is composed of an activation function, which first merges the vectors of three elements (feature, mutator, and parameter) through a fully connected layer, and then passes through the nonlinear function softmax:

\begin{equation}
f_{att}=softmax(FCnet(\mathcal{U})+FCnet(v_{m})+ FCnet(v_{p}))
\end{equation}

Combining all $\alpha$ together, we get the vector $\mathcal{A}_{1 \times N} = \langle\alpha_{1},\alpha_{2},\cdots,$\\$\alpha_{N}\rangle$. We normalize $\mathcal{A}$ to matrix $\mathcal{W}= \langle \alpha'_{11},\alpha'_{12},\cdots,\alpha'_{1N}\rangle$ and expand the dimension of $\mathcal{W}$ to $D'\times N$ to form a mask matrix $\mathcal{M}$, which allows $\mathcal{M}$ to be applied to each vector in $\mathcal{U}$.

\begin{equation}
\mathcal{W}=\frac{\mathcal{A}}{\Sigma\{\alpha_{i}|\alpha_{i}\in\mathcal{A}\}}
\end{equation}

\begin{equation}
\mathcal{M} = 
\begin{bmatrix}
	\mathcal{W}\\
	\mathcal{W}\\
	\vdots\\
	\mathcal{W}
\end{bmatrix}
_{D' \times N}
\end{equation}

We scale the elements in the matrix $\mathcal{M}$ to $\mathcal{U}$ to get the final attention layer output vector $\mathcal{U'}$. Next, we use a fully connected layer for classification. Through training the following attention layer, we can further visualize the information from feature $\mathcal{U}$ through a heat map, which measures the importance of the input features.
In order to get the heat maps, we cluster the data set according to different mutators. For each cluster, we feed the data into the attention model, and obtain the matrix $\mathcal{A}$ from each input while calculating the accuracy. Finally, we obtain each mutator using a heat map, shown in Figure~\ref{model}. The darker red positions represent the corresponding seed areas which play a more important role in the classification and vice versa.

In summary, our attention model is mainly composed of three parts. First, according to whether the input type is a picture or a byte sequence, we use three layers of CNN or RNN to extract features respectively. Second, \tool introduce an attention layer described above to apply attention weight to each feature and finally we pass the weighted features to a fully connected layer for classification. 

% \sj{the example should be expanded with details here; for instance, what are the uncovered nodes; what are the probabilities; and what are the result of the RNN model. Make it intuitively clear.}

\begin{example}
Take the program in Figure~\ref{abstraction} for example, after 1 hour's fuzzing, AFL can only cover line 2, 3, 4, 6 with the given seeds. So line 7 becomes the uncovered block of the program. Line 6 is the critical block (only 1/1000 inputs reach line 6).
% We observe that only one in ten thousand input really reached line 6, which is the critical block of line 7. 
AFL spends most of its effort on test cases which fail to reach line 6, i.e., 90\% of the generated test cases cannot arrive at line 6.
% Thus, even if AFL chooses a seed file that can set the flag to 1, the input coverage will be destroyed due to random mutation.
\tool introduces an attention model to predict whether a certain mutation on a certain input can cover line 6.
% classify the input while extracting the influence of different mutators on different positions of the seed file. 
% After the model is trained, 
For instance, the influence of mutator $Arth_{-}$ on the seed file \{<a=200, b=-1, c=-10, buf[0]="Y", buf[1]="Y", buf[2]="Y" > can be visualized as the heat map at the bottom of Figure~\ref{model}. We could see that variables $a$ and $b$ are important to the coverage with very high weight, and the $Arth_{-}$ operation on other variables does not affect the coverage of program execution. This is consistent with the logic of the program. For the $>$ and $==$ operations corresponding to variables $a$ and $b$, reducing the number of them will likely lead to a failure of the comparison, and the coverage of the corresponding program will also change. Similarly, for $Arth_{+}$, variables $b$ and $c$ are important to the coverage of the seed file, and adding variable $a$ will not affect the result of comparison $>$.
\end{example}

%Take input \{<a=200, b=-1, c=-100, buf[0]="Y", buf[1]="Y", buf[2]="Y" >, $Arth_{-}$, 10; line 6\} for program in Figure~\ref{mot} as en example. As shown in the Figure~\ref{model}, we apply weights for each feature extracted by attention so we can learn that for the $Arth_{-}$ operation, the $a$ and $b$ variables in the seed file are closely related to the coverage (their weights are very high), and the $Arth_{-}$ operation on other variables will not affect the coverage of program execution. 

\subsection{Mutation Guidance}
\label{mg}

% Once the fuzzing process encounters a bottleneck, the carrier fuzzer often spends a lot of energy on paths that have been executed and do not have many basic blocks afterward (such as error handling modules). In order to expand the fuzzing to uncovered blocks, we try to guide the carrier fuzzer to generate more inputs that can reach the pre-dominant block. After reward filtering, we have a pre-dominant block list of the key blocks. we apply deep learning to each of these blocks. 
With the attention model, we can not only learn whether an input can reach a specific block, but also get the heat maps for each mutator. \tool then uses the heat maps to 
% With the help of the heat map, we can 
guide the fuzzer to generate inputs that can reach the selected critical blocks.
% so that the fuzzing process is focused on solving the bottleneck. 
In the heat map, \emph{the level of heat quantifies the importance of the corresponding position of a seed file in the classification result under a certain mutator}. The hotter it is, the more important it is in the classification, which means those `hot bytes' (with heat values larger than a threshold) determine whether an input can reach the critical blocks of interest. If the hot bytes are mutated with the specific mutation operator, there is a high probability that the coverage will be changed. 
% pre-dominant block cannot be reached.

%\begin{figure}[h]
%  \centering
%  \includegraphics[width=\linewidth]{mutation-guidance.pdf}
%  \caption{Mutation guidance workflow}
%  \label{mutation}
%\end{figure}

\begin{algorithm}[t]
    Let $seed$ be the seed file\;
	Let $B_{critical}$ be the set of blocks selected by reward\;
	Let $mutator\_list$ be the set of mutator\;
% 	Let $heatmap\_list$ be the set of heat map\;
% 	Let $loc$ be the target mutation location\;
	
$seed\_cov = GetCoverage(seed)$\;
% \If{$(b_c\_list = seed\_cov \cap B_{critical})\neq \emptyset$} {
    \For{$mutator \in mutator\_list$} { 
    let $threshold$ be the average value of $seed$'s heat map of $mutator$\;
    	\For{each byte $b$ in $seed$} {
    	   % \For{$b_c \in b_c\_list$} {
    	        let $heat$ be the value in $b$'s heat value\;
    	        
        % 	}
        		\If{$
        % 		\exists(
        		heat > threshold
        % 		)
        		$} {
        			skip this mutation with a certain probability $p$\;
        			continue\;
        	}
        	Apply $mutator$ on $seed$\;
    	}
    }    
% }					

	\caption{Mutation Guidance}
	\label{alg3}
\end{algorithm}

Algorithm~\ref{alg3} shows the details of how \tool efficiently generates a large amount of inputs that can reach the critical blocks with the guidance of the heat maps. We first obtain the seed file coverage (line 4). If the seed file can cover any of the critical blocks that we select, then for each mutator, we skip the hot bytes (determined at line 9) to avoid change in the coverage (line 10). We remark that a seed file may cover multiple critical blocks. In this case, as long as a byte is not a hot byte for all the blocks, it is mutated. Besides, \tool still adopts some randomness to improve the diversity of mutation, i.e., choosing to mutate the hot bytes with a small probability $p$.    
% If the seed file can cover multiple pre-dominant blocks, as long as there is a heat map that meets the conditions, then we still mutate the corresponding positions. 
% It makes it possible to generate more diversified inputs on the premise that the pre-dominant block can still be reached to break through the bottleneck. 
% However, the further improvement of coverage may also be related to bytes with high heat value. Simply skipping bytes will directly reduce the input space, and there may still be no new coverage breakthroughs after a round of mutation. Therefore, we skip the bytes with high heat value with a certain probability to ensure that the input is generated efficiently while keeping the input space unchanged. For those seed files that fail to cover any pre-dominant block we care about, \tool uses the original fuzz strategy and does not use heat maps to guide mutation behavior.

\begin{example}
Following the example in Section~\ref{tam}, the seed file can cover line 6 and we introduce our mutation guidance strategy.
% to mutation. 
% It is worth noting that the heat map is diverse for different mutators. 
Take $Arth_{-}$ as an example whose heat map is shown in Figure~\ref{attoverview}. We set the `hot' threshold in Algorithm \ref{covguided} as the average of all the heat values of the seed file, i.e., 0.4794 in this example. Variables $a$, $b$ both have heat values that are greater than the threshold. So, for the $Arth_{-}$ operation, we avoid the subtraction of $a$ and $b$ and mutate the remaining variables with high probability. Similarly, for the $Arth_{+}$ operation, we avoid adding variables $b$ and $c$ because their heats are great than the average heat 0.5123.
\end{example}

\subsection{Overall Algorithm}

\begin{algorithm}[t]
	\caption{Main Algorithm}
	\label{main_alg}
	Let $\mathcal{P}$ be the target program\;
    Let $\mathcal{B}$ be the set of bugs\;
    Let $iter\_limit$ be the maximum iterations\;
    $CFG = StaticAnalysis(\mathcal{P})$\;
    
    \For{$seed\in seed\_pool$}{
        Let $length$ be the length of $seed$\;
        Let $seed\_coverage$ be the coverage of executing $\mathcal{P}$ using $seed$\;
        \For{$iterations \leq iter\_limit$} {
            \If{EncounterBottleneck} {
                $uncovered\_blocks$ = $FindUncov(CFG, cov)$\;
                $rewards$ = $RewardCal(CFG, cov)$\;
                $B_{critical}$ = $Select(uncov\_blocks, rewards, CFG)$\;
                \For{$block\in B_{critical}$} {
                   $heatmap\_list$ = $TrainModel(dataset)$\;
                }
            }
            
            \For{$mutator \in mutator\_list$} {
                \For{$loc \leq length$} {
                    \If{EncounterBottleneck} {
                        $Guide(heatmap\_list, mutator, loc)$\;
                    }
                    $input$ = $mutate(seed, mutator, parameter)$\;
                    $cov\_result$ = $Excute(\mathcal{P}, input)$\;
                    $dataset$ = $Record(seed, mutation, cov\_result)$\;
                    % $coverage$ = $UpdateCov($\;
                    \If{$result == Crash$} {
                        $\mathcal{B}.append(input)$\;
                    }
                    \If{$HasNewCov(cov\_result)$} {
                        $seed\_pool.append(input)$\;
                    }
                }
            }
        }
    }
    \Return $\mathcal{B}$\;
\end{algorithm}

Putting all together, we summarize our overall algorithm in Algorithm~\ref{main_alg}. A mutation budget for each seed is set as $iter\_limit$. We first obtain the CFG of the target program $\mathcal{P}$ at line 4. Then, for each seed in the seed pool, within the mutation budget (line 8), we first determine whether a bottleneck is met at line 9 (details explained later in Section \ref{sec:id}). Note that this is often not the case in the initial phase of fuzzing. So,  
% normally at the beginning of fuzzing, since the coverage of the program has risen rapidly and the bottleneck we defined is not encountered, 
\tool will initially execute the carrier fuzzer according to the default strategy (line 15-25). 
% (line 4, 5, 6, 14, 15, 18, 19, 22, 23, 24, 25 in Algorithm~\ref{main_alg}). 
During the process, \tool mutates the seed (line 19), executes the program (line 20) and collects the data (line 21). If an input incurs new edge coverage (line 24), it will be added into the seed pool (line 25). After a while, a bottleneck might be met (line 9), this is when \tool starts to work by finding the uncovered blocks (line 10), evaluating the rewards of covering them (line 11) and select the critical blocks from their pre-dominant blocks (line 12). Then for each critical block, we train an attention model for each mutator using the dataset collected (line 14). After the models are trained, \tool goes on to guide the subsequent fuzzing process at line 18. Note that whenever a crash is triggered (line 22), we add the input to the bug-triggering inputs $\mathcal{B}$ (line 23). Different from existing learning-enabled fuzzing ~\cite{godefroid2017learn, haller2013dowsing, wang2017skyfire, bastani2017synthesizing, sivakorn2017hvlearn, sivakorn2017hvlearn, byrd1995limited, nichols2017faster}, \tool is more effective in accurately locating the bottleneck and paying better attention to those valuable bytes and mutators to break them. 
% firstly shows which program blocks deserve the fuzzer’s attention in fuzzing. Secondly, instead of filtering invalid inputs, we directly change the mutation strategy to generate high-quality input so that fuzzer's attention is focused on the bottleneck.
% records more fine-grained information (seed file, mutation, coverage) to prepare for the next steps (line 3, 20, 21 in Algorithm~\ref{main_alg}).
% After a period of time, the coverage of the program tends to grow slowly and fuzzing encounters bottleneck (line 8 in Algorithm~\ref{main_alg}).
% In order to concentrate the attention of fuzzing to break through the bottleneck, we first calculate the reward for all uncovered nodes through the coverage information collected before and static analysis of the program, thereby filtering out the critical blocks that play a key role in improving the program coverage (line 9, 10, 11 in Algorithm~\ref{main_alg}).
% For each critical block selected, we train the deep learning model with attention mechanism through the previously collected data. While the model classifies the input, it extracts input heat maps which represents the correlation between the input byte and the coverage (line 12, 13 in Algorithm~\ref{main_alg}).
% Guided by heat maps, we skip the mutation of the hot byte in seed files, so that rich input can be generated while ensuring the coverage of the seed file remains unchanged (line 16, 17 in Algorithm~\ref{main_alg}).

% Next

% \sj{discuss the algorithm - walk it through with the example; briefly discuss the key difference from existing ML-based fuzzing algorithms.}
~\begin{example}
Take the program in Figure~\ref{abstraction} as an example. At the beginning, \tool runs AFL with its default mutation strategy and collect the data. In the early stage of fuzzing, AFL was able to successfully cover line 2 to 6. But after 1 hour, AFL meets the bottleneck. 
% In our definition, fuzzing encountered a bottleneck (line 8 in Algorithm~\ref{main_alg}). 
We identify the uncovered blocks, which is the line 7. By reward calculation and computing the pre-dominant block, we select line 6 as the critical block for learning. \tool uses the previously collected data to train the attention model and obtain the heat maps. 
% While classifying whether the input obtained after mutation of a seed file can reach line 6, we can get heat maps through the attention model. 
% Based on the above information, we 
\tool prioritizes seed files that can reach line 6 and further guides the mutation according to the heat map so that a large number of inputs can be generated setting $flag$ to 1. The strategy enables \tool to break the bottleneck efficiently and reach line 7 (triggering the bug) within 10 minutes (while AFL fails in over 24 hours).
\end{example}

\section{Implementation Details}\label{sec:id}
% In this section, we introduce the implementation details of \tool in each phase.
\vspace{1mm}
\noindent{\textbf{Data Collection}}
In the initial stage, \tool runs the carrier fuzzer with the default configuration. As mentioned before, the data we collect are seed files, mutators, mutation parameters and the basic block coverage of each program execution. We use \textsc{AFL++}~\cite{fioraldi2020afl++} as our carrier fuzzer. Note that compared with AFL, AFL++ provides accurate coverage information (without hash collision) as well as the above-mentioned additional information which are required by our approach.

Unfortunately, in terms of detailed coverage information, a fuzzer like AFL, which uses edges as coverage statistics, does not provide accurate coverage information. Instead of comprehensively recording the complete execution paths, AFL uses a compact hash bitmap to store code coverage. This compact bitmap is highly efficient but not informative enough for our purpose. When the program is executed, AFL can only know whether a new edge is triggered or not but does not have any idea of the specific position of the edge in the program's CFG and due to hash collision, it is infeasible to identify which edge is covered. There are some mature instrumentation methods, such as LLVM's sanitizer coverage to obtain the coverage of each execution. However, we found that these methods have a noticeable overhead for program execution, which leads to a significant decrease in the fuzzing speed. To solve this problem, we try to only use the information that AFL provides.  

% To understand how we obtain program coverage, we first explain how AFL records the program coverage. AFL instruments the program's basic blocks and uses the hash between two basic blocks to represent an edge. The edge hash calculation algorithm of AFL is as follows:

% \begin{center}
% {\small \begin{verbatim}
%             cur_location = <RANDOM>;
%             shared_mem[cur_location ^ prev_location]++;
%             prev_location = cur_location >> 1;
% \end{verbatim}}
% \end{center}

% When the source code is available, $cur\_location$ represents the random number inserted by AFL when instrumenting the current block. Otherwise, $cur\_location$ represents the address of the basic block in the binary. \emph{shared\_mem} is the shared memory used by AFL to record hash. $prev\_location$ represents the random number or address of the previous basic block. The hashing algorithm is a basic $XOR$ operation. Each time the program is executed, AFL obtains the hash value of each edge covered by the input according to the above algorithm. 
% The $raw\_cov$ in the algorithm is the hash value we recorded.

\tool extracts the ICFG of the program by statically analyzing the program's instrumented binary file and calculates the hash for each edge in the same way as AFL. When fuzzing the program, whenever AFL gets the hash of the program execution path, we can look up the dictionary and get its corresponding basic block. Based on the dictionary, we can further obtain the exact program coverage from the existing records of AFL with little overhead. 
%As AFL knows the number of each block and the execution hash at runtime, we only need to build a dictionary to identify the relationship between the hash value and the basic block number. 
We remark that for large programs, it is inevitable that the edges will have hash collisions (while for small programs, this problem is negligible). In this case, a hash value may correspond to multiple edges, making the coverage of each execution biased. This problem has been discussed in~\cite{gan2018collafl},
% For large programs, this problem of AFL has also been discussed. [Collafl] 
which proposes a new hash calculation method to mitigate path collisions while trying to preserve low instrumentation overhead. Note that \tool is flexible to incorporate such method to obtain more accurate coverage information. 

% In particular, we used static analysis described above to build a hash dictionary. For the binary file using AFL instrumentation, we use disassembly tools like \textsc{objdump} to extract the random number corresponding to each basic block. For ordinary binary files, we also simulate AFL and mark basic blocks by extracting each of their address. Then, we construct the ICFG of the program through the static assembly program written on our own. Note that we supplement the missing indirect calls in static analysis through dynamic fuzzing data.
% For each edge of the program, we followed the AFL method for hash calculation. So we can get a hash value for each edge between two basic blocks, and this hash value and the two blocks form the key-value pair of the dictionary. During the execution of AFL, we only record the hash value to save the time spent searching for basic blocks and leave the work of looking up the table to the reward calculation phase.

\vspace{1mm}
\noindent\textbf{{Heat Map Acquisition}}
In practice, the coverage of critical blocks tends to be polarized, i.e., they are either covered by almost every input, or they are only covered by few. To solve this problem, we under-sample the data to ensure it has a balance distribution. 
% Through the reward calculation stage, we get a series of critical blocks. 
Thanks to the fast speed of fuzzing, although the probability of covering certain interesting blocks is often relatively low, 
% when fuzzing encounters a bottleneck, 
we are able to collect millions of fuzzing data, which is enough for us to train a reasonable model.
% including program inputs and there corresponding coverage 
We randomly select the same amount of positive and negative labeled data to keep the training data as balanced as possible.

We train deep learning models using \textsc{PyTorch} (version 1.6.0). For feature extraction, we use 3 layers CNN or RNN for picture or byte sequence respectively. We generate a mask at the attention layer by performing non-linear operation $softmax$ on the mutator and parameter and then adding them together. We use a fully connected layer for classification after multiplying the normalized mask by the corresponding pixel/byte. We use cross-entropy loss function in our model and Adam optimizer~\cite{da2014method} with learning rate 0.001 to help the model converge quickly. The attention model is trained for 60 epochs to achieve high accuracy (about 85\% on average). 
After model training, we cluster the data set according to the mutator. By feeding the data into the trained model and calculating the attention mask, we can generate each mutator's heat map for different seed files. Note that it takes 40 to 60 minutes to train the model and obtain the heat map on average, which could be paralleled with the normal fuzzing process.
% Due to the introduction of the attention layer, the model will take about 40 -60 minutes to train. 

%Note that training will bring a certain overhead, but first, fuzzing mainly uses CPU to run, and model training is more dependent on GPU, they basically do not compete with each other for computing resources, and secondly, while model training, fuzzing can still use the default strategy to generate inputs without having to pause and wait.

\vspace{1mm}
\noindent\textbf{{Mutation Guidance}}
% Using heat map for mutation guidance is the key to improving the efficiency of fuzzing. 
As mentioned in section \ref{mg}, we choose not to mutate the hot bytes with high probability. 
% if the heat is very high, 
% it means that this location is significant for whether the input can reach a certain critical block. We choose to avoid mutating these pixels/bytes when fuzzing. And what is the appropriate heat threshold?
We use the average heat value to determine whether a byte is hot (larger than the average heat). 
% thi threshold to ensure that we can accurately select threshold for each mutator. 
% As mentioned before, skipping all these bytes will lead to a reduction in the input space. 
Meanwhile, we choose to mutate these hot bytes with a probability of 5\% to introduce some randomness in the mutation.

\vspace{1mm}
\noindent\textbf{{Bottleneck Judgement}} Intuitively, fuzzing encounters a bottleneck if the coverage does not increase over certain amount of time. In practice, we check whether a bottleneck is met every hour 
(which aligns to the model training time) 
by calculating if the coverage increase in the last hour is smaller than a threshold, in our case, 5\%. 

\section{Evaluation}\label{sec:ev}
%In this section, we evaluate the performance of \tool from the perspectives of code coverage and the number of bugs found and discuss how our algorithm specifically helped the fuzzing process.

In this section, we evaluated \tool from two aspects. First, we compare the performance of \tool with state-of-the-art fuzzers including: AFL, AFLfast~\cite{8233151}, Vuzzer~\cite{rawat2017vuzzer}, Driller~\cite{stephens2016driller} and NEUZZ~\cite{she2019neuzz} to assess its effectiveness and efficiency.
AFLfast performs seed file scheduling through statistics of execution path to optimize AFL.
Vuzzer uses taint analysis to determine the relevant bytes in the input to improve the mutation efficiency.
Driller as the state-of-the-art hybrid fuzzer, solve the path constraint by symbolic execution to improve coverage.
NEUZZ introduces neural network to smoothing programs, helping fuzzer build inputs efficiently.
These Baseline fuzzers have its own advantages in some respects, which can reflect \tool's effectiveness in seed file and mutation scheduling as a hybrid fuzzer.
Then, we evaluate the usefulness of each component of \tool. 

Specifically, we aim to answer the following research questions through our experiments.
\footnote{\tool is available at https://github.com/ICSE-ATTuzz/ATTuzz}

\begin{itemize}
    \item[\emph{RQ1:}] \emph{Does \tool improve code coverage effectively?}
    \item[\emph{RQ2:}] \emph{Does \tool allow us to break the bottleneck effectively?}
    \item[\emph{RQ3:}] \emph{Are reward computation and attention guidance complementary to each other?}
    \item[\emph{RQ4:}] \emph{Does \tool allow us to discover more bugs?}

\end{itemize}
~\\
\textit{\textbf{Experiment Settings}}

\vspace{1mm}
\textbf{Target Programs:} We evaluate \tool on two different types programs: (i) 9 real-world programs, as shown in Table~\ref{eva}, (ii) LAVA-M dataset \cite{dolan2016lava}. To demonstrate the performance of \tool, we compare 1) the edge coverage and 2) the number of bugs detected.
%by \tool to various kinds of state-of-the-art fuzzers, including AFL-style fuzzers AFL \cite{AFL}, AFLFast \cite{8233151}, hybrid fuzzer Vuzzer \cite{rawat2017vuzzer} and ML-based fuzzer NEUZZ \cite{she2019neuzz}.

\begin{table}[]
\begin{tabular}{|c|c|c|c|}
\hline
\multicolumn{2}{|c|}{\textbf{Programs}}                                          & \multirow{2}{*}{\textbf{Size}}                                      & \multirow{2}{*}{\textbf{Version}} \\ \cline{1-2}
Class & name                                                                     &                                                                     &                                   \\ \hline
JPEG  & libjpeg                                                                  & 1.1M                                                                & 9c                                \\ \hline
TTF   & harfbuzz                                                                 & 6.1M                                                                & 2.0.0                             \\ \hline
PDF   & mupdf                                                                    & 45.8M                                                               & 1.12.0                            \\ \hline
XML   & libxml                                                                   & 9.2M                                                                & 20907                             \\ \hline
ZIP   & zlib                                                                     & 441k                                                                & 1.01b                             \\ \hline
BIN   & \begin{tabular}[c]{@{}c@{}}readelf\\ objdump\\ size\\ strip\end{tabular} & \begin{tabular}[c]{@{}c@{}}3.4M\\ 13.1M\\ 9.2M\\ 10.7M\end{tabular} & 2.30                              \\ \hline
\end{tabular}
\vspace{1mm}
\caption{Programs in the experiments.}
\label{eva}
\end{table}

\vspace{1mm}
\textbf{Experimental Setup:} The host used in the experiment has 16-core CPU (Intel R core (TM) i9-9900k CPU @ 3.60GHz), 32G GPU (Tesla V100-SXM2) and 64G main memory. Note that each fuzzing task only takes one CPU core. 
We run each fuzzer for the same time budget with the same initial seed corpus and compare their edge coverage achieved and the number of bugs found. We compare the average coverage results of different fuzzers given the same time budget: 24 hours with 5 times.
% To investigate the coverage performance, we compare the fuzzers on fixed runtime budget. 
Since NEUZZ needs to run AFL for an hour to generate the initial seed corpus in advance, for a fair comparison, we also spend 1 hour in advance to learn the initial heat map when the first bottleneck is encountered. For other fuzzers, we add the seed files obtained within the first hour to the initial seed file pool in advance.
~\\

% \vspace{1mm}
\textit{\textbf{RQ1. Does \tool improve code coverage effectively?}}

\begin{figure*}[t]
  \centering
  \includegraphics[width=.95\linewidth]{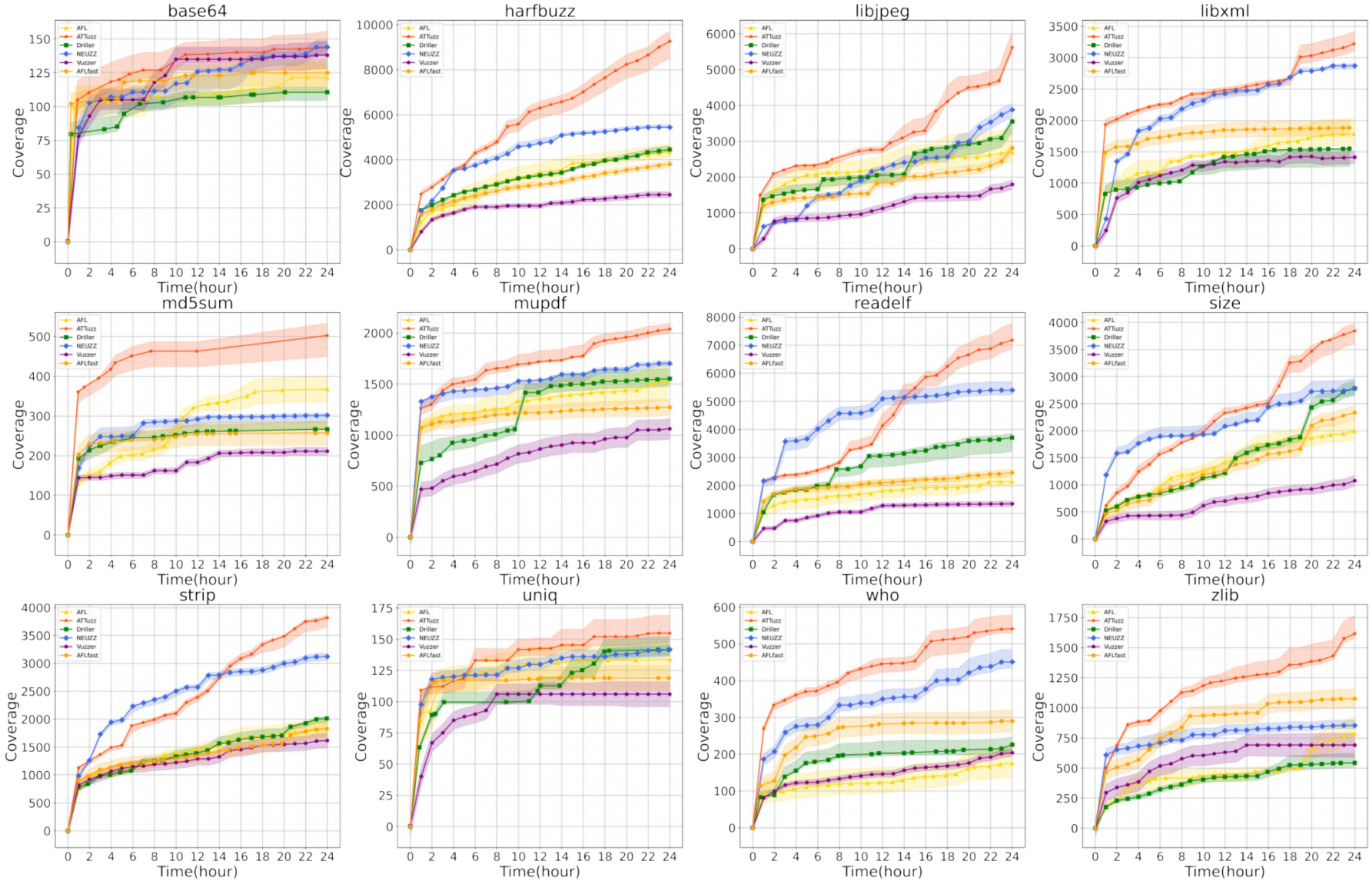}
  \caption{Coverage growth through 24 fuzzing.}
  \label{expcov}
\end{figure*}

We first discuss the overall coverage improvement
% In order to evaluate performance in program coverage, 
% AFL, AFLfast, and NEUZZ are all based on AFL, and the plotdata file in the output directory stores detailed information about the coverage.
% This evaluation shows not only 
from two aspects: 1) the total number of new edges found (Table~\ref{comp}) and 2) the rate of improving the edge coverage over time (Figure~\ref{expcov}). 
% We collect the edge coverage information from AFL’s edge coverage report. 
% The results are summarized in Table~\ref{comp} and Figure~\ref{expcov}.
Our experimental settings are consistent with the recommendation from ~\cite{klees2018evaluating}.
As shown in Table~\ref{comp}, the solid lines represent the average data of the 5 times repeated fuzzing data and the shadows represents the confidence intervals of 95\%.
For all programs, \tool outperforms  AFL, Driler, Vuzzer and AFLfast in terms of overall coverage, and outperforms NEUZZ in all programs. On average, \tool achieves 100\% more coverage than AFL, 90\% more than Driller and 20\% more than NEUZZ. 
Furthermore, \emph{for programs including base64, libjpeg, md5sum, mupdf, and readelf, \tool achieves the same or even more coverage in the first two hours than that of AFL, AFLfast and Driller in 24 hours}. This shows the superior performance of \tool in improving program coverage than state-of-the-art approaches. A closer look reveals that these programs share a common feature, that is, theirs inputs are highly structured, e.g., the JPEG format or the ELF format, which are suitable for deep learning with attentions to extract useful features and select the right bytes and mutators for mutation. Besides, we could observe from Figure \ref{expcov} that the baseline fuzzers tend to get stuck in the bottleneck after some period of fuzzing while \tool can consistently improve coverage in the long run, which is clearly evidenced in programs including harfbuzz, libjpeg, libxml, mupdf, readelf, size, strip, who and zlib. We further run libjpeg using \tool and AFL for a long time, i.e., 5 days to show that \tool is consistently improving coverage. The result shows that \tool achieves about 50\% more coverage than AFL in the end (5026 V.S. 3297).  
% Whether it is the JPEG image format handled by libjpeg or the ELF format of readelf, they have clear specifications to handle data in a special format.

% In general, \tool can achieve significantly higher edge coverage compared to other gray-box fuzzers like AFL and is comparable to the latest method of introducing deep learning.

\begin{table}[]
\resizebox{\linewidth}{!}{
\begin{tabular}{cccccll}
\hline
\textbf{Programs} & \tool & \textbf{NEUZZ} & \textbf{AFL} & \textbf{Driller} & \textbf{Vuzzer} & \textbf{AFLfast} \\ \hline
libjpeg           & \textbf{5260}        & 3879           & 2696         & 3555             & 1794            & 2812             \\
harfbuzz          & \textbf{9268}        & 5440           & 4344         & 4838             & 2449            & 3810             \\
mupdf             & \textbf{2039}        & 1701           & 1520         & 1694             & 1062            & 1274             \\
libxml            & \textbf{3221}        & 2870           & 1786         & 2097             & 1413            & 1883             \\
zlib              & \textbf{1617}        & 854            & 786          & 543              & 691             & 224              \\
readelf           & \textbf{7534}        & 5398           & 2142         & 4140             & 1346            & 2461             \\
size              & \textbf{3847}        & 2848           & 1995         & 2848             & 1077            & 2336             \\
strip             & \textbf{3818}        & 3122           & 1852         & 2685             & 1617            & 1828             \\
base64            & \textbf{144}         & \textbf{144}   & 111          & 121              & 138             & 125              \\
md5sum            & \textbf{512}         & 319            & 368          & 328              & 211             & 257              \\
who               & \textbf{525}         & 461            & 176          & 226              & 204             & 290              \\
uniq              & \textbf{155}         & 142            & 134          & 140              & 106             & 119              \\ \hline
\end{tabular}
}
\caption{Comparing edge coverage for 24 hours' runs.}
\label{comp}
\end{table}

~\\
\vspace{1mm}
\textit{\textbf{RQ2. Does \tool allow us to break the bottleneck effectively?}}

\begin{figure}[t]
  \centering
  \includegraphics[width=\linewidth]{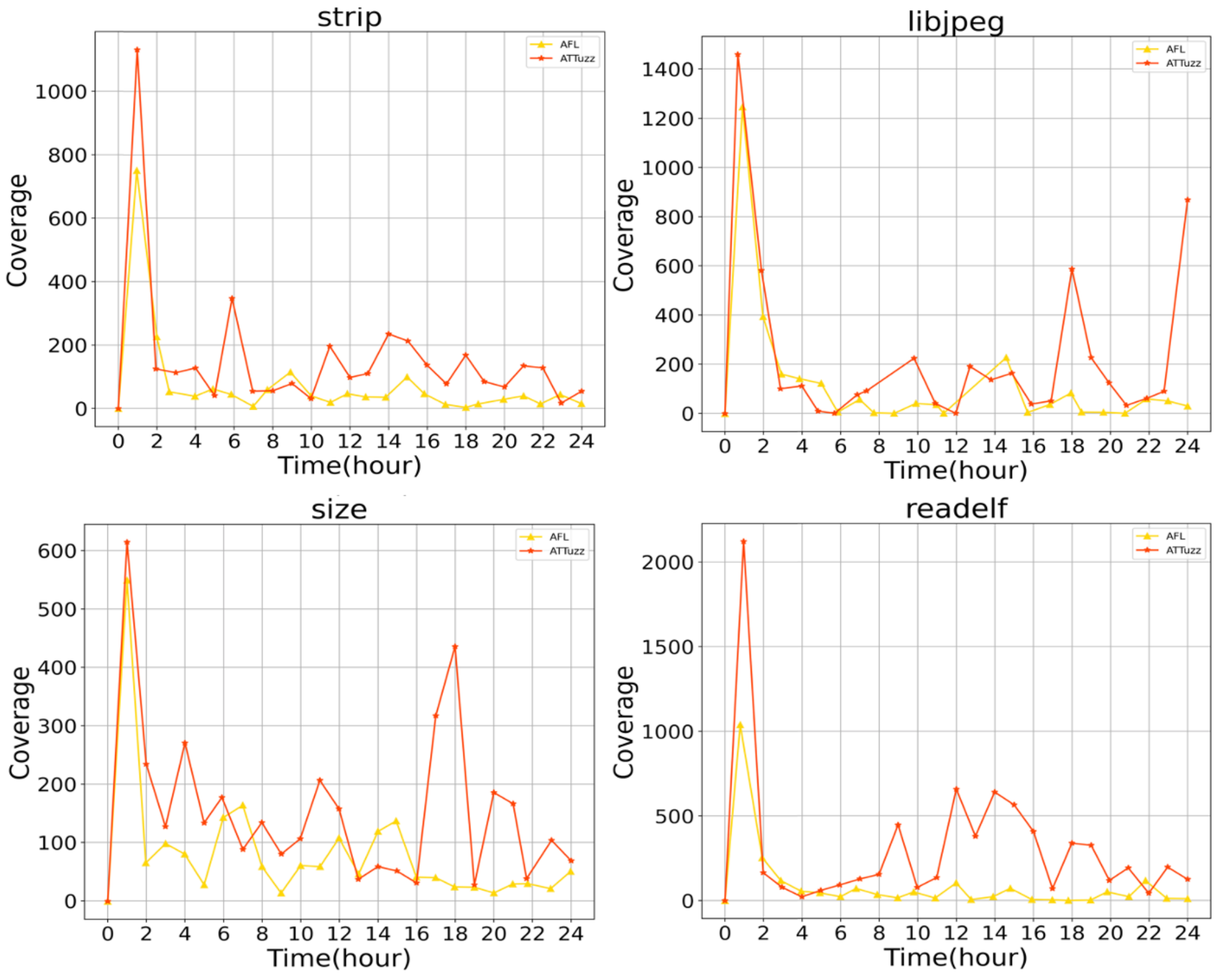}
  \caption{The normalized derivative average value.}
  \label{bottleneck}
\end{figure}

We further discuss the trend of the coverage improvement to show that existing fuzzers get stuck after a while whereas \tool is able to break the bottlenecks during the fuzzing process in a measurable way as follows. 
% ours keeps improving.
We calculated the coverage growth rate in each hour and show the results in Figure~\ref{bottleneck} (we only show the comparison of AFL and \tool for the sake of space\footnote{Comparisons on the other programs share a similar trend and are available at https://github.com/ICSE2022/ATTuzz}). 
% As shown in Figure~\ref{bottleneck}, 
We could observe that at the beginning, both fuzzers do not encounter any bottlenecks and have a high coverage growth rate. However, in the later stage of fuzzing, AFL struggles to achieve new coverage, i.e., having a significantly lower coverage growth than \tool. Thanks to reward calculation and attention guidance, \tool is able to repeatedly break the bottlenecks (even in the very late hours especially for libjpeg and size) and achieve the highest coverage in 24 hours compared to baseline fuzzers. More comprehensive results can be found at our github site, which consistently shows the effectiveness of \tool in breaking the bottlenecks.

\begin{figure}[t]
  \centering
  \includegraphics[width=.98\linewidth]{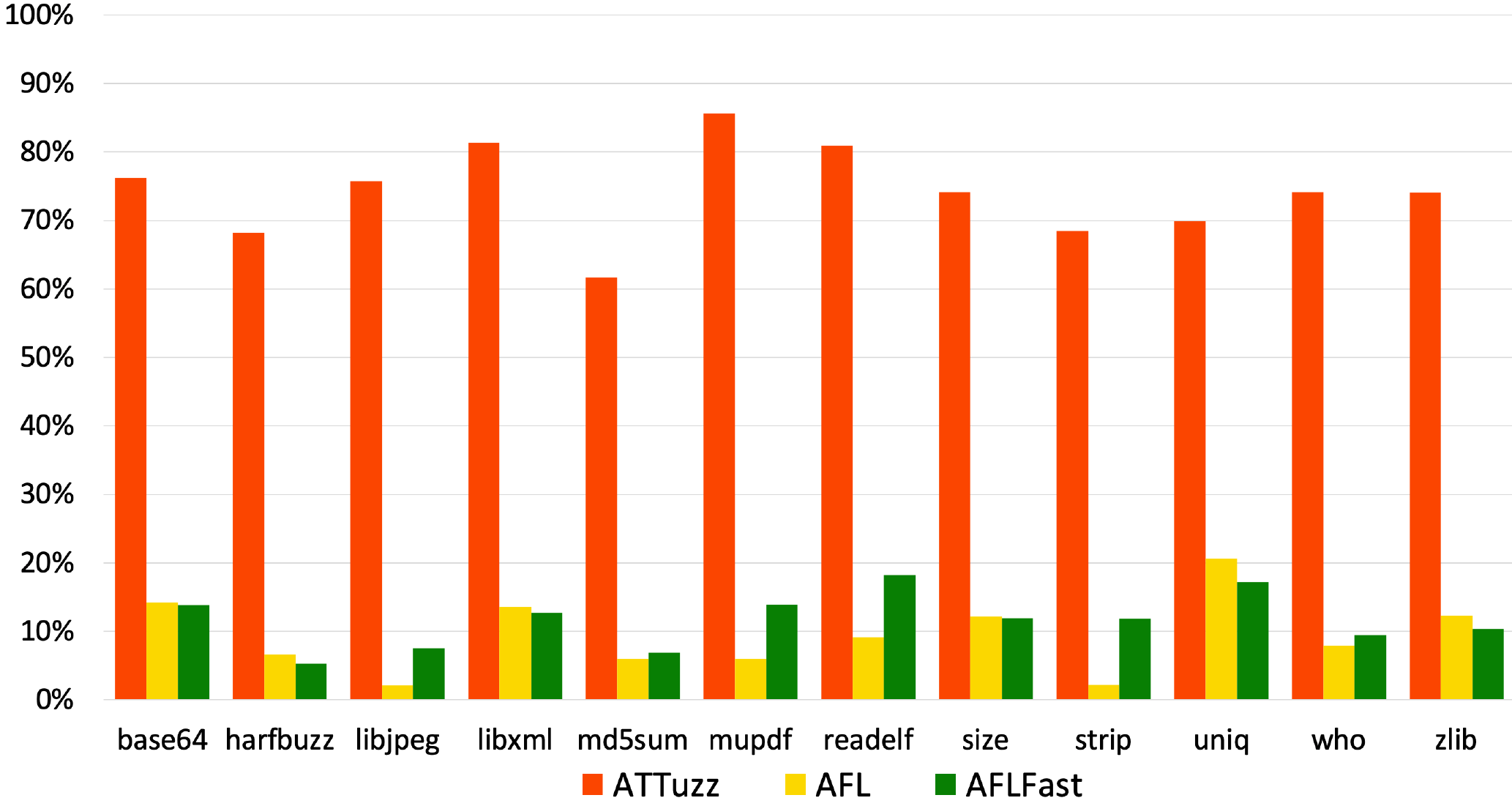}
  \caption{Critical block coverage ratio.}
  \label{ratio}
\end{figure}

\textbf{Evaluation of mutation guidance.} Recall that the heat maps are used by \tool to guide the mutation on the right bytes of the input with right mutators to better reach the critical blocks and further break the bottleneck. 
% Through the heat map, we extracted the relationship between the different location in the input and the program coverage, and we generated the input that can reach the critical block by guiding mutation. 
We use the ratio of inputs to trigger the critical blocks by different fuzzers 
% when mutation guidance is used to visually illustrate 
to show the effectiveness of our heat map based mutation guidance. The results are shown in Figure \ref{ratio}. 
% For each program, we observe the ratio of the input generated that can reach the target critical block when heat-map-based mutation guidance is performed, and finally calculate the average value. Figure~\ref{ratio} shows the corresponding coverage ratio results for the program we selected. 
We have performed mutation ratio statistics on the AFL-based Fuzzer. We observe that the input generated by AFL and AFLfast has a very sparse coverage ratio of the critical blocks, i.e., only about 10\% of the inputs can cover them. In other words, most fuzzing effort is spent in program branches that are less useful to induce new coverage. 
Note that to keep certain randomness (in the same spirit of MCMC~\cite{andrieu2003introduction}) in the exploration, \tool also mutates the hot bytes with a certain small probability (5\% in our experiment and flexible to adjust). 
% which leads to our ratio statistic results to be compromised, but overall, 
Overall, guided by the heat maps, \tool pays good attention to those critical blocks, i.e., about 75\% of the input generated by \tool can reach the target critical blocks, which greatly increases the chance of breaking the bottlenecks.

\begin{figure}[t]
  \centering
  \includegraphics[width=\linewidth]{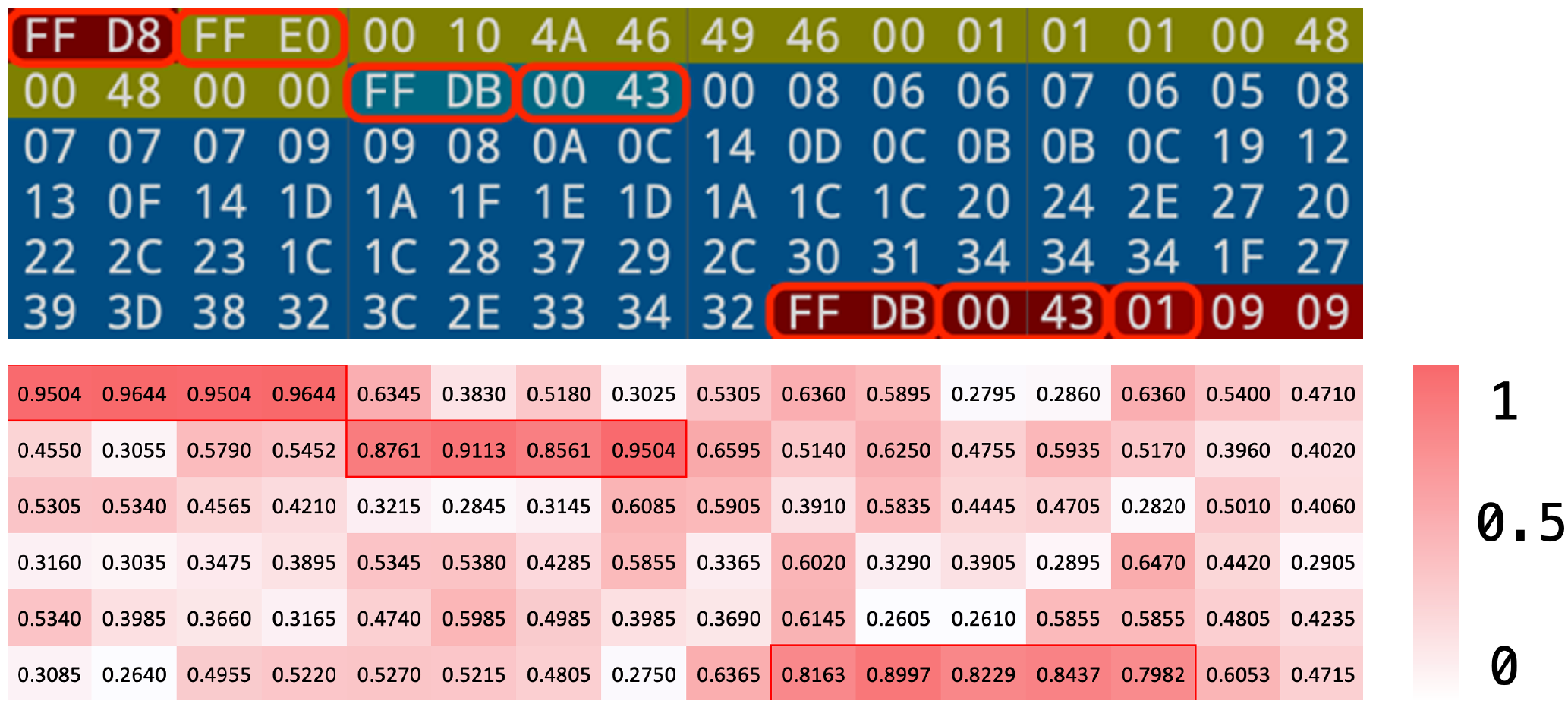}
  \caption{A fragment of the JPEG file format.}
  \label{jpeg}
\end{figure}

\begin{figure}[t]
  \centering
  \includegraphics[width=\linewidth]{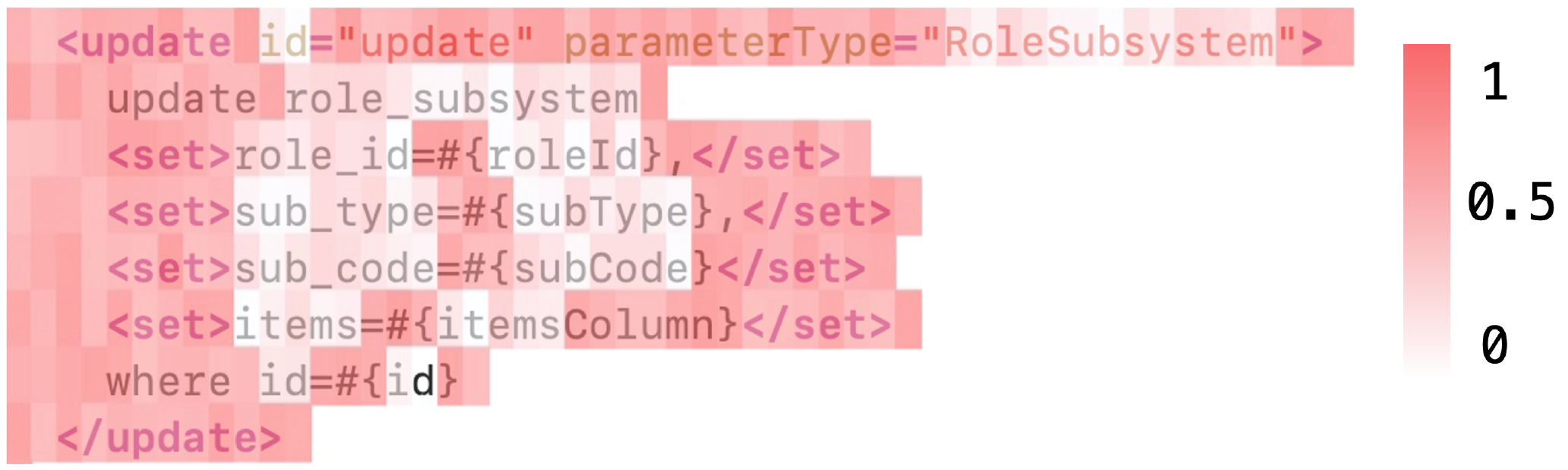}
  \caption{A fragment of the xml file format.}
  \label{xml}
\end{figure}

\textbf{Case study of heat map.} Take libjpeg as an example. The upper part of Figure~\ref{jpeg} is a beginning fragment of a jpeg format file opened in \textsc{010editor}, where different colors represent different areas and structures of the file. Specifically, the parts in the red circle represent a special field identifier. If any of these bytes is changed, the file cannot be recognized as a jpeg file normally, while mutating the rest bytes will not directly lead to such a problem. 
The correspondence between the program coverage and byte mutation is successfully reflected in the heat map (the lower part of Figure ~\ref{jpeg}), i.e., the bytes in the red circles mostly have high heat values (hot bytes). \tool avoids mutating these bytes to ensure that the generated inputs are still valid. Similarly, we observe that bytes that are critical for the validity of XML syntax are successfully identified as hot-bytes. Hot bytes in Figure ~\ref{xml} indicate those words which are closely related to the xml syntax, and mutating them will directly cause the xml file fail to execute (invalid inputs).

~\\
\vspace{1mm}
\textit{\textbf{RQ3. Are reward computation and attention guidance complementary to each other?}}

\begin{figure}[t]
  \centering
  \includegraphics[width=\linewidth]{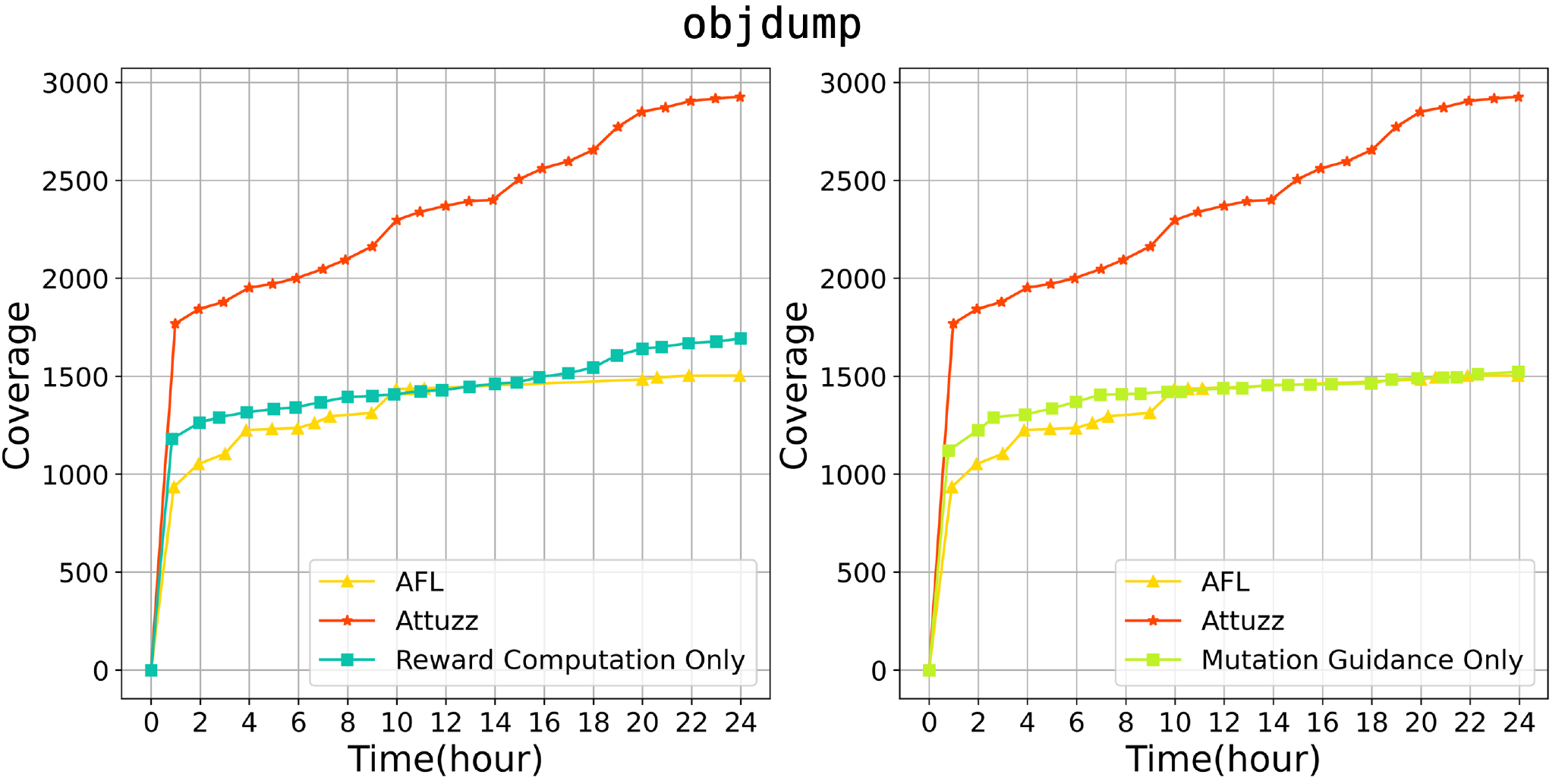}
  \caption{Coverage result with individual component.}
  \label{individual}
\end{figure}

In the following, we show that either techniques by itself is not impressive using two sets of experiments. 
% For reward computation, 
First, we keep using reward calculation to guide fuzzer to select seed files that can reach the critical blocks, but we do not use the attention models to guide mutations on these seed files to show the effectiveness of the mutation guidance. Second, we omit reward calculation and only randomly select k\% of all uncovered blocks and train attention model on their pre-dominant blocks. We conduct mutation guidance towards covering these blocks to show the effectiveness of reward calculation. The results are shown in Figure~\ref{individual}. The figure on the left and right shows the result of the first and second set of experiments respectively. We observe from the left figure that selecting the valuable seed file only has a slight improvement in the performance of fuzzing (compared to AFL). And the effect of randomly selecting blocks for mutation guidance is almost negligible as evidenced by the right figure. In summary, the results show that either technique by itself is not impressive to improve the performance of fuzzing while combining them will significantly boost fuzzing's performance. The probable reason is that even if the seed file is selected correctly, random mutation will destroy the effectiveness of the input, while random selection of blocks will lead to a high probability of selecting non-rewarding blocks.
(the generated input would have a high probability of triggering this node, or this node has little child nodes).
Similar results are obtained on other programs as well.

~\\
\vspace{11mm}
\textit{\textbf{RQ4. Does \tool discover more bugs?}}
% We show new bugs we found ...

The main purpose of fuzzing is to find as many bugs as possible. In this experiment, we evaluate each fuzzer's capability of detecting bugs in the software binaries. We summarize the results in Table~\ref{bug}. Note that we omit those software that all fuzzers fail to find any bug. For the remaining programs, \tool is able to find 9X, 7X, 4X, 4X and 1.5X times of bugs than Vuzzer, Driller, AFLFast, AFL and NEUZZ respectively, which shows the usefulness of improved coverage by \tool in discovering more bugs. Besides, \tool can not only find the maximum number of bugs, but also trigger more kinds of bugs than other fuzzers in baseline.
%We show a case study in of a bug discovered only by \tool to further show the effectiveness in appendix.
% is able to locate 4 times bugs than AFL in total. Compared to NEUZZ, we can also find 40\% more bugs.

\begin{table}[]
\resizebox{\linewidth}{!}{
\begin{tabular}{ccccccc}
\hline
\textbf{Programs} & \textbf{\tool} & \textbf{NEUZZ} & \textbf{AFL} & \textbf{AFLfast} & \textbf{Driller} & \textbf{Vuzzer} \\ \hline
\multicolumn{7}{c}{Bug Detection}                                                                                         \\ \hline
mupdf             & \textbf{8}    & 0              & 0            & 0                & 0                & 0               \\
readelf           & 29            & 16             & 3            & 5                & 7                & 5               \\
objdump           & \textbf{8}    & \textbf{8}     & 6            & 6                & 2                & 0               \\
size              & 2             & \textbf{6}     & 0            & 4                & 0                & 1               \\
strip             & \textbf{20}   & \textbf{20}    & 7            & 5                & 2                & 2               \\ \hline
\multicolumn{7}{c}{Bug type}                                                                                              \\ \hline
interger overflow & $\checkmark$  & $\checkmark$   & $\times$     & $\times$         & $\times$         & $\checkmark$    \\
heap overflow     & $\checkmark$  & $\checkmark$   & $\times$     & $\checkmark$     & $\times$         & $\times$        \\
invalid pointer   & $\checkmark$  & $\times$       & $\checkmark$ & $\checkmark$     & $\checkmark$     & $\times$        \\
out-of-memory     & $\checkmark$  & $\checkmark$   & $\checkmark$ & $\checkmark$     & $\checkmark$     & $\checkmark$    \\
assertion crash   & $\checkmark$  & $\checkmark$   & $\times$     & $\times$         & $\times$         & $\times$        \\ \hline
Total             & \textbf{67}   & 50             & 16           & 20               & 10               & 8               \\ \hline
\end{tabular}
}
\caption{Bug Detection.}
\label{bug}
\end{table}

\section{Conclusion}\label{sec:con}
We present \tool, an efficient fuzzer which pays better attention on 1) the valuable blocks identified using a reward calculation mechanism based on the fuzzing data, and 2) the valuable bytes and mutators using a deep learning network with attention mechanism to focus on those critical seed files and mutations. We further demonstrate how attention models with heat maps can be utilized to guide more effective test inputs generation to help the fuzzer break the bottlenecks met. We extensively evaluate \tool on 13 real-world programs comparing to 5 state-of-the-art fuzzers of 3 different categories, and show that \tool achieves over 50\% more coverage and detects up to 11X bugs. \tool shows the potential of paying attentions with the help of deep learning whilst fuzzing and extends the possibility of detecting more vulnerabilities by fuzzing with attentions.  
% Our results demonstrate 
% that by focusing the fuzzer's attention on the off position through the reward and attention model, we can greatly improve the efficiency of fuzzing.

% \section*{Acknowledgments}
% \section{Appendices}
% \newpage
\bibliographystyle{plain}
\bibliography{references}
% \newpage
\section{Supplementary Material}\label{sec:sm}

\textbf{A Motivating Example}

% \todo{It would make a much better impression if the example can be some real bug - even better some CVE.}In this section, we use a simple program (shown in Figure~\ref{mot}) to illustrate the motivation of our work. 
% Figure~\ref{motivating} shows a concrete code snippet of our motivating example.The code segment itself is very simple. 
We start with a motivating example shown in Figure~\ref{mot}. First, the program checks a nested condition determined by three variables $a$, $b$, and $c$. The $Flag$ is set to 1 if the nested condition is satisfied. Then, the program determines if the corresponding bytes in the buffer are equal to $X$. If yes, a bug is triggered.
% When all the conditions are met, we can reach the real functional module, where bugs often hide.

% What is the problem if we use AFL to fuzz such a program fragment? 
Such a simple program brings several challenges to coverage-guided fuzzers like AFL. First, \emph{the nested conditions and checksum are not easy to be satisfied by random mutations (Challenge 1),} i.e., the probability to generate a seed covering new control locations is extremely low. Second, even if we finally meet the conditions and obtain the corresponding seed, \emph{the effectiveness of the seed will be easily destroyed by the random mutation mechanisms (Challenge 2).} 
For example, after a period of fuzzing, AFL finally met the conditions of line 2, 3 and 4 in Figure~\ref{mot}, and obtained a seed file that can set $Flag$ to 1. In order to further improve coverage, AFL mutates this seed to generate more inputs. Unfortunately, such random mutations are hardly useful as they very likely set flag to false. For instance, AFL fails to cover line 8 after 12 hours and 13 million runs even a seed which could set $Flag$ to 1 is given.  
% \sj{Provide some statistics here to support the argument - like AFL failed to cover the branch in 2 hours ... }
% For AFL, even after giving a seed file that can set Flag to 1, after 12h fuzzing and 13 million runs, it cannot reach 8 lines. First, AFL still uses the default strategy to select seed files from the massive seed file pool. As a result, only 8\% of the seed files selected by AFL can set the flag to 1 after the seed files reaching line 7 are successfully generated. What's more, the random mutation strategy results in that even if the correct seed file is selected, only 1.3\% of the newly generated input can still reach line 7

% For example, a simple $bitflip$ can change the value from a positive number to a negative number, causing the condition in the line 4 of the Figure~\ref{motivating} to not be met. Even if we successfully meet the condition of the 7 row of the function, we still cannot reach the target basic block because we have destroyed the condition of the flag. In response to this problem, we proposed our strategy to make the mutation efficient through guidance.

To tackle the above challenges, different works have been proposed which can be roughly divided into two categories.
% To solve this problem, the methods proposed by the existing work can be roughly divided into two categories. 
The first category used program analysis techniques such as dynamic taint analysis~\cite{rawat2017vuzzer,chen2018angora}, concolic execution~\cite{stephens2016driller,yun2018qsym,cadar2008klee,king1976symbolic} and static analysis~\cite{li2017steelix, rawat2017vuzzer} to identify the most relevant bytes to mutate (a.k.a. solving the magic byte problem) directly. While program analysis based approaches can accurately obtain the value of the magic byte (with a powerful solver at hand), the subsequent mutation may still destroy such ``magic''. The other category introduced machine learning to filter less meaningful inputs (potentially with no coverage gain)~\cite{zong2020fuzzguard} or use it to generate more valuable inputs~\cite{rajpal2017not, she2019neuzz}. However, existing machine learning based approaches have limited effectiveness in selecting control locations that can maximize the coverage reward. For instance, Neuzz's gradient guided mutation strategy can only keep 10\% of newly generated inputs reaching line 7 with seeds setting $Flag$ to 1. Besides, the adopted machine learning models (e.g., LSTM~\cite{rajpal2017not} and NN~\cite{she2019neuzz}) are often considered lacking of interpretability.
% % 
% % \sj{similarly, provide some statistics here to support the argument - like neFuzz failed to cover ...}

% % Similarly Neuzz uses a seed selection strategy similar to that of AFL, so although it has gradient-guided mutation, only about 10\% of newly generated inputs can set flag to 1 after reach reaching line 7.
 
% % still cause the input to fail to execute to the target code segment.   
% % [RNNfuzzer][Neuzz].
% % It is true that the method of program analysis can accurately obtain the specific value of the magic byte, but as mentioned earlier, even if the magic byte is satisfied, the mutation strategy may still cause the input to fail to execute to the target code segment.
% % [RNNfuzzer] does not give a method for selecting the nodes that need training, while [Neuzz] learns the overall CFG of the program. The method they proposed reduces the possibility of destroying the previous conditions to a certain extent, but makes the input control flow lacks directivity.

% % Our method can break through the key basic blocks that have not yet been covered without destroying the prerequisites, so as to improve the overall coverage of the program more efficiently.

In this work, we aim to address both challenges systematically by deepening our understanding from two sides, i.e., the program and the fuzzing process, and their interactions. First, we use a carrier fuzzer (e.g., AFL) to generate a number of tests and run the target program with them. Note that the program is instrumented to collect the execution trace of each input, which contains more fine-grained information than coverage. After a while, \tool determines that covering some of the blocks that are yet to be covered will be more rewarding than others by calculating a \emph{coverage reward} for each uncovered control locations based on the abstraction of the program estimated from the fuzzing data. In the running example, line 7 will have the highest reward. 
%as it will trigger the buggy code. 
\tool then selects the target control locations for machine learning by prioritizing those with high rewards. In the machine learning phase, \tool adopts state-of-the-art deep learning models with attention mechanism, which allows us not only to predict whether a specific combination of seed and mutator can reach the target location, but also to obtain the ``explanation'' (in form of heat maps) which informs us the importance of different bytes (under a mutator) on the program coverage (e.g., can reach line 7 or not). With the explanation, \tool then guides the fuzzer to mutate the corresponding position of the input with different probabilities. In the example, if the seed input can already set $Flag$ to 1, we will avoid performing $bitflip$ on $b$, and allow $arth$ to operate on $c$. In this way, the effectiveness of the seeds will be preserved and many more inputs satisfying the nested conditions in line 2, 3 and 4 will be generated, thereby increasing the probability of triggering the bug in line 8.

\begin{figure}[t]
{\footnotesize
\begin{verbatim}
            void func(int a, int b, int c, char* buf) {
            1.    int Flag = 0;
            2.    if(a > 100) {
            3.        if(b == -1) {
            4.            if(c < 0) {
            5.    	          Flag = 1;
                             return 0;
                          }
                      }
                  } 
            6.    if(Flag) {
            7.        if(buf[0] == "X") {
            8.            buggy code ...;		
                      }    	
                  }
            9.    return 1;
              }
\end{verbatim}}
\caption{A motivating example}
\label{mot}
\end{figure}
% breaking the bottleneck of row 9 instead of generating a large number of invalid inputs based on seeds. 
% Finally, in the updating phrase, XXX performs the above steps for newly encountered uncovered blocks to maximize the entire program's coverage. Below we provide the details of these modules.  

% In the model training phase, in addition to simple classification, XXX adds an attention mechanism, which allows us not only to predict whether the combination of seed and mutation can trigger the target block but also to obtain \textbf{heat maps} which tells us that performing different mutations to the variable $a$, $b$, $c$, and $buf$ will affect whether the program can execute to line 7. 

% In the \textbf{mutation strategy updating phrase}, according to the heat maps of different mutators, XXX guides the fuzzer to mutate the corresponding position of the input with different probabilities. For example, for the fragment in figure~\ref{motivating}, if the seed file can already set the flag to one, we will avoid performing $bitflip$ operations on the variable $b$, and allow $arth-$ to operate on the variable $c$. In this way, we can generate rich input while ensuring that the conditions of rows 2, 3, and 4 in figure~\ref{motivating} are met, thereby breaking the bottleneck of row 9 instead of generating a large number of invalid inputs based on seeds. 
% Finally, in the updating phrase, XXX performs the above steps for newly encountered uncovered blocks to maximize the entire program's coverage. Below we provide the details of these modules.

~\\
\textbf{Derivative comparison}

Table~\ref{Normalized} shows the normalized average derivative through 24 fuzzing. The greater the derivative, the more efficient the fuzzy can break through the bottleneck and increase the program coverage. Since \tool continue to break through the bottleneck of in the later stage of fuzzing, its average derivative is nearly 1.5 times that of the second place.

\begin{table}[H]
\resizebox{\linewidth}{!}{
\begin{tabular}{ccccccc}
\hline
\textbf{Programs} & \textbf{ATTuzz} & \textbf{NEUZZ} & \textbf{AFL} & \textbf{Driller} & \textbf{Vuzzer} & \textbf{AFLfast} \\ \hline
base64            & \textbf{0.263}  & 0.194          & 0.135        & 0.178            & 0.09            & 0.141            \\
harfbuzz          & \textbf{0.307}  & 0.18           & 0.144        & 0.16             & 0.081           & 0.126            \\
libjpeg           & \textbf{0.219}  & 0.183          & 0.164        & 0.182            & 0.114           & 0.137            \\
libxml            & \textbf{0.243}  & 0.216          & 0.135        & 0.158            & 0.106           & 0.142            \\
md5sum            & \textbf{0.343}  & 0.181          & 0.167        & 0.115            & 0.147           & 0.048            \\
mupdf             & \textbf{0.327}  & 0.234          & 0.093        & 0.18             & 0.058           & 0.107            \\
readelf           & \textbf{0.257}  & 0.19           & 0.133        & 0.19             & 0.072           & 0.156            \\
size              & \textbf{0.256}  & 0.209          & 0.124        & 0.18             & 0.108           & 0.123            \\
strip             & \textbf{0.188}  & 0.180          & 0.142        & 0.155            & 0.176           & 0.16             \\
uniq              & \textbf{0.257}  & 0.16           & 0.184        & 0.164            & 0.106           & 0.129            \\
who               & \textbf{0.279}  & 0.245          & 0.094        & 0.12             & 0.108           & 0.154            \\
zlib              & \textbf{0.195}  & 0.178          & 0.168        & 0.176            & 0.133           & 0.149            \\ \hline
\end{tabular}
}
\caption{Normalized derivative through 24h's fuzzing}
\label{Normalized}
\end{table}
~\\
\textbf{Case Study}

\begin{figure}[H]
  \centering
  \includegraphics[width=.8\linewidth]{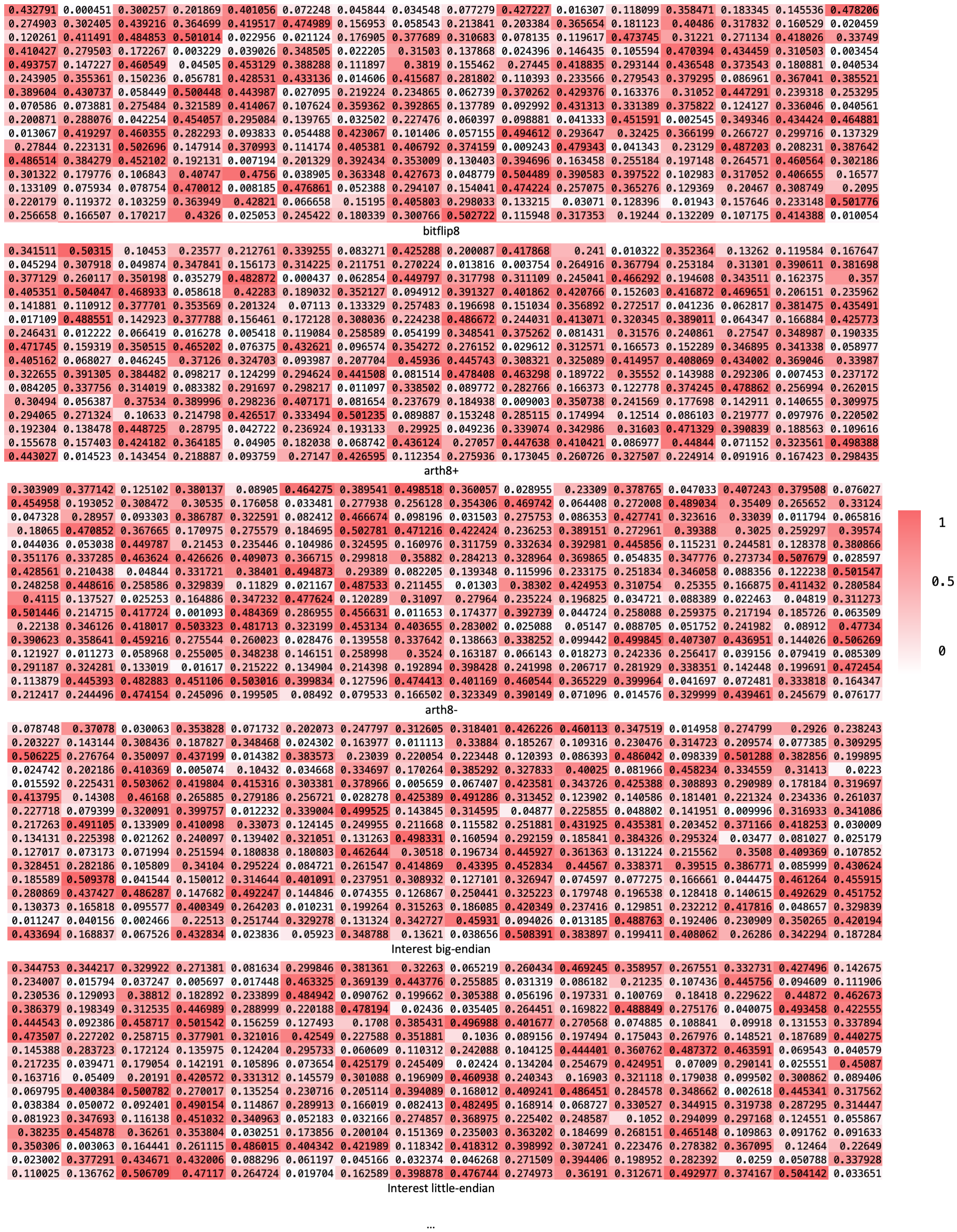}
  \caption{some of the heat maps.}
  \label{case}
\end{figure}

We explain the effectiveness of ATTuzz by triggering BUG in Figure~\ref{code} in mupdf which is only triggered by ATTuzz in our experiment. As shown in Figure~\ref{code}, after a period of fuzzing, AFLpp can cover lines 1-5 without the help of ATTuzz but it was unable to cover line 6-7. In this case, ATTuzz intervenes to help Fuzzing improve overall coverage. First, ATTuzz builds CFG by static analysis of the program, and uses the dynamic data generated by Fuzzing to supplement the CFG, and then the edge probability of the program Markov chain is estimated. According to the reward mechanism, ATTuzz calculates the reward of basic blocks and finds the line 6 has a high Reward (8.214 in this case). By tracing pre-dominant block, ATTuzz considers line 5 as a critical block, and integrate the collected fuzzing data to prepare for the next learning stage. Thanks to the attention mechanism, ATTuzz's model can learn the heat maps of different mutators under different seed files while classifying the inputs. Figure~\ref{case} shows some of the heat maps. When selecting seed files, we first filter out those that can reach line 5, so as to increase the possibility that the generated input can still reach the critical block. When mutating the seed file, we skip the mutation of high heat bytes according to different mutators and corresponding heat maps, so as to ensure that more input can be generated on the premise of reaching line 5. Under the guidance of ATTuzz, AFLpp triggered the bug of line 7 and further explored other branches of the program.

\begin{figure}[H]
{\footnotesize
\begin{verbatim}
1.  fz_append_display_node(
        fz_context *ctx,
        fz_device *dev,
        fz_display_command cmd,
        int flags,
        const fz_rect *rect,
        const fz_path *path,
        const float *color,
        fz_colorspace *colorspace,
        const float *alpha,
        const fz_matrix *ctm,
        const fz_stroke_state *stroke,
        const void *private_data,
        int private_data_len)
    {
    ...
2.      if (list->len + size > list->max)
    	{
    		...
3.     		list->list = fz_realloc_array(  //allocated here
       		    ctx, list->list, 
       		    newsize, 
       		    fz_display_node); 
4.      list->max = newsize;
            ...
    	}
        ...
5.   	if (private_off)
    	{
6.  	    char *out_private = (char *)(void *)(&node_ptr[private_off]);
7.      memcpy(out_private, private_data, private_data_len); 
        	//overflow
    	}
8.    	list->len += size;
    } 
\end{verbatim}}
\caption{A code fragment in mupdf 1.12.0}
\label{code}
\end{figure}

\end{document}